\documentclass{article}

\usepackage{arxiv}

\usepackage[utf8]{inputenc} 
\usepackage[T1]{fontenc}    
\usepackage{hyperref}       
\usepackage{url}            
\usepackage{booktabs}       
\usepackage{amsfonts}       
\usepackage{nicefrac}       
\usepackage{microtype}      
\usepackage{lipsum}		
\usepackage{graphicx}

\usepackage{dcolumn}
\usepackage{bm}
\usepackage{amsmath}
\usepackage{amssymb}
\usepackage{amscd}
\usepackage{amsthm}
\usepackage{amsfonts}
\usepackage{graphicx}
\usepackage{multirow}

\newtheorem{theorem}{Theorem}

\newtheorem{condition}[theorem]{Condition}

\newtheorem{definition}[theorem]{Definition}

\newtheorem{proposition}[theorem]{Proposition}
\newtheorem{remark}[theorem]{Remark}

\date{\today}

\author{
Jos\'{e} M. Amig\'{o} \\
Centro de Investigaci\'{o}n Operativa, \\
Universidad Miguel Hern\'{a}ndez, \\
Elche, 03202 Alicante, Spain \\
\texttt{jm.amigo@umh.es}
\And
Roberto Dale \\
Centro de Investigaci\'{o}n Operativa, \\
Universidad Miguel Hern\'{a}ndez, \\
Elche, 03202 Alicante, Spain \\
\texttt{rdale@umh.es}
\And
Juan C. King \\
Centro de Investigaci\'{o}n Operativa, \\
Universidad Miguel Hern\'{a}ndez, \\
Elche, 03202 Alicante, Spain \\
\texttt{juan.king@goumh.umh.es}
\And
Klaus Lehnertz \\
Department of Epileptology, \\
University of Bonn Medical Centre, \\
Venusberg Campus 1, \\
53127 Bonn, Germany \\
Helmholtz Institute for Radiation and Nuclear Physics, \\
University of Bonn, Nussallee 14--16, \\
53115 Bonn, Germany \\
\texttt{klaus.lehnertz@ukbonn.de}
}
\renewcommand{\headeright}{}
\renewcommand{\undertitle}{}

\begin{document}

\title{Generalized synchronization in the presence of dynamical noise and its detection via recurrent neural networks}
\maketitle

\begin{abstract}
 Given two unidirectionally coupled nonlinear systems, we
speak of generalized synchronization when the responder \textquotedblleft
follows\textquotedblright\ the driver. Mathematically, this situation is
implemented by a map from the driver state space to the responder state
space termed the synchronization map. In nonlinear times series analysis,
the framework of the present work, the existence of the synchronization map
amounts to the invertibility of the so-called cross map, which is a
continuous map that exists in the reconstructed state spaces for typical
time-delay embeddings. The cross map plays a central role in some techniques
to detect functional dependencies between time series. In this paper, we
study the changes in the \textquotedblleft noiseless
scenario\textquotedblright\ just described when noise is present in the
driver, a more realistic situation that we call the \textquotedblleft noisy
scenario\textquotedblright . Noise will be modeled using a family of driving
dynamics indexed by a finite number of parameters, which is sufficiently
general for practical purposes. In this approach, it turns out that the
cross and synchronization maps can be extended to the noisy scenario as
families of maps that depend on the noise parameters, and only for
\textquotedblleft generic\textquotedblright\ driver states in the case of
the cross map. To reveal generalized synchronization in both the noiseless
and noisy scenarios, we check the existence of synchronization maps of
higher periods (introduced in this paper) using recurrent neural networks
and predictability. The results obtained with synthetic and real world data
demonstrate the capability of our method.
\end{abstract}

\tableofcontents

\begin{quotation}
{\bf The first description of synchronization of two coupled dynamical
systems (two pendulum clocks hanging from a beam) is attributed to
Christiaan Huygens in 1665. In 1990, Pecora and Carroll demonstrated that
chaotic systems that are unidirectionally coupled (i.e., drive-response
systems) can also synchronize. By synchronization in both of the previous
cases we mean that the two systems evolve in finite time to a dynamic with a
constant relationship between their states. In turn, chaotic synchronization
gave rise to generalized synchronization, where now the relationship between
states may be arbitrary. Precisely, our work deals with a mathematical
formulation of generalized synchronization in the more realistic case of
drivers perturbed by dynamical noise. In addition, we do not assume
knowledge of the states but only scalar observables of them in the form of
time series. We also discuss other practical issues, most importantly a
method to detect generalized synchronization for both noiseless and noisy
drivers, based on recurrent neural networks. The capability of this method
is successfully tested with numerical simulations and real world data.}
\end{quotation}

\section{Introduction}

The framework of this paper is nonlinear time series analysis, and the topic
is synchronization of two unidirectionally coupled nonlinear systems and its
generalization when noise is present in the driving system. By
synchronization we mean \textit{generalized} (or \textit{general}) \textit{%
synchronization}\ in the sense of Afraimovich et al. \cite{Afraimovich1986}
and Rulkov et al. \cite{Rulkov1995}, i.e., there is a map, called the 
\textit{synchronization map}, that transforms the states of the driving
system (\textit{driver}) into states of the driven system (\textit{responder}%
), possibly with a time delay or after an initial transient time. Identical
(complete, full, ...) synchronization corresponds then to the
synchronization map being the identity between two structurally equal
systems; other, more interesting examples include lag, intermittent-lag and
phase synchronizations \cite{Rosenblum1997}. Synchronization, whether
identical or generalized, plays an important role in many fields of science
and engineering, particularly in nonlinear dynamics \cite%
{Pecora1990,Boccaletti2002}, telecommunications \cite%
{Pecora1990,Cuomo1993,Kocarev1995}, neuroscience \cite%
{Nowotny2008,Amigo2015,Hovel2020}, and cryptography \cite%
{Kocarev1992,Amigo2007,Kinzel2010,Banerjee2011}; see, e.g., Pikovsky et al. \cite{Pikovsky2001}, and Pecora and Carroll \cite{Pecora2015} for overviews and historical notes.

In the noiseless\ or fully deterministic scenario, synchronization has been
extensively studied using a number of techniques, including cross (or
mutual) prediction \cite{Schiff1996,LeVanQuyen1999}, conditional Lyapunov
exponents \cite{Pecora1990,Hunt1997,Pyragas1997}, replica synchronization 
\cite{Abarbanel1996}, asymptotic stability of the responder \cite%
{Kocarev1996}, nonlinear interdependence measures \cite%
{Arnhold1999,QuianQuiroga2000,QuianQuiroga2002}, cellular nonlinear networks 
\cite{Sowa2005,Krug2007}, complexity measures extracted from symbolic
representations \cite{Monetti2009,Amigo2012,Amigo2013}, reservoir computing 
\cite{IbanezSoria2018,Lymburn2019}, and more. We will use prediction because
predictability is a fingerprint of determinism, i.e., functional dependence 
\cite{Kantz2004}.

Furthermore, it is known \cite{LeVanQuyen1999} that, in the case of two
unidirectionally coupled nonlinear systems with a noiseless driver, there
exists typically a continuous map defined from the reconstructed state space
of the responder to the reconstructed state space of the driver, that was
called the closeness mapping in Amig\'{o} and Hirata \cite{Amigo2018} and will be called the 
\textit{cross map }here. As it turns out, the definition of synchronization
amounts to the invertibility (i.e., bijectivity) of the cross map; in fact,
the inverse of the cross map is the \textquotedblleft
translation\textquotedblright\ of the synchronization map (if any) from the
original domains (driver and responder state spaces) to the reconstructed
ones. The existence of the cross map has been used to study interdependence
and causal relationships in nonlinear time series analysis \cite%
{Sugihara2012,Harnack2017,Amigo2018,Ying2022}. In a nutshell, these methods
harness some actual or hypothetical property of the cross map (continuity,
smoothness, local expansiveness) to reveal, given bivariate time series of a
coupled dynamics, what the driving system is. We will generalize the cross
and synchronization maps to multi-time versions that are well suited to the
application of recurrent neural networks in time series analysis.

The main objective of this paper is the extension of the cross and
synchronization maps from noiseless to noisy drivers. To model noise in the
driver we replace the dynamic of a noiseless driver with a family of driving
dynamics indexed by a finite number of parameters whose values are randomly
chosen, an idea called finitely parameterized stochasticity \cite%
{Muldoon1998}. To implement this idea in our setting, we will use the 
\textit{stochastic forcing} approach of Stark et al. \cite{Stark2003}.
First, noise is formulated as an autonomous dynamical system called a 
\textit{shift system}, whose states comprise all possible noise realizations
in form of parametric sequences; the $n$th component of a given sequence
indicates which is the chosen driving dynamic at time $n$. Second, the noisy
driver is then formulated as a non-autonomous system, namely, a system
forced by that shift system. As a result, our approach to synchronization in
the presence of dynamical noise is based on state space reconstruction for
unidirectionally coupled systems \cite{Stark1999} and stochastic forcing 
\cite{Stark2003}, and is sufficiently general for practical purposes. We
will show that the cross and synchronization maps can be extended from the
noiseless to the noisy scenario by incorporating an additional dependence on
the noise parameters and only for typical driver states in the case of the
cross map.

Beside discussing theoretical results in the noiseless and noisy scenarios,
we also explore synchronization with synthetic and real world data. Prompted
by multi-time expressions for the synchronization map, we not only use 
\textit{perceptrons} but also \textit{long short-term memory} (LSTM) \textit{%
nets} \cite{Goodfellow2016}. In this approach, synchronization is detected
via estimation of a responder state by a contemporaneous driver state or, in
case of LSTM\ nets, by a segment of contemporaneous and past driver states.
As the benchmark in numerical simulations we chose \textit{nearest-neighbor} 
\textit{cross prediction} because it is based precisely on the existence of
the cross map, so it fits very well in our approach. In fact, the continuity
of the cross map entails that near neighbors of a point in the responder
state space map to near neighbors of its image in the driver state space
(and vice versa when synchronization sets in). Therefore, one expects that
this correspondence stays if low-amplitude noise affects the driving
dynamic. We also apply LSTM nets to detect coupling directionality and
synchronization in electroencephalograms (EEGs) from a subject with
epilepsy. The optimization of the parameters and metaparameters of our
numerical tools is beyond the scope of the present work.

To address the points described above, the rest of this paper is divided in
a first, theoretical part and a second, numerical part. Thus, in Section \ref%
{sec2} we first review the basics of our approach to make this paper
self-contained. For didactic reasons, we start with the Takens and Stark (or
forced Takens) embedding theorems (Section \ref{sec2.1}), along with the
concept of cross map (Section \ref{sec2.2}); then we introduce the concept
of generalized synchronization (Section \ref{sec2.3}) and discuss its
relationship with the cross map (Section \ref{sec2.4}). Novel concepts such
as the cross and synchronization maps of higher periods are introduced for
further applications in Sections \ref{sec6} and \ref{sec7}. The presentation
is rigorous from a mathematical point of view, but unnecessary
technicalities are avoided. Along the way, practical issues are also
considered with a view to the second part of the paper. Once the
traditional, noiseless scenario has been presented, the noisy scenario is
set in two steps: in Section \ref{sec3} we revisit stochastic forcing and an
embedding theorem that is used in the second step, Section \ref{sec4}, where
a unidirectional coupling with a noisy driver is modeled as stochastic
forcing. The generalization of the cross and synchronization maps to the
noisy scenario is the subject of Section \ref{sec5}. The theoretical
concepts of Sections \ref{sec2}-\ref{sec5} are illustrated and put into
practice in the second part of the paper. For this purpose, in Section \ref%
{sec6} we resort to two unidirectionally coupled H\'{e}non maps,
synchronization being detected with recurrent neural networks, and compare
the results with those obtained with nearest-neighbor cross predictability.
In Section \ref{sec7} we tackle the applicability of our tools to the
analysis of real data in the form of EEGs, where noise and bidirectional
coupling are the rule. In this case, the \textquotedblleft
driver\textquotedblright\ is identified by the direction with the strongest
coupling. Our findings are discussed in light of results published in the
specialized literature. Finally, the main contributions and conclusions of
this paper are summarized in Section \ref{sec8}.


\section{The noiseless scenario}

\label{sec2}

This section is a compact, mathematically oriented account of the cross map,
synchronization and their interplay in the absence of noise.

\subsection{Embedding theorems}

\label{sec2.1}

Following Stark \cite{Stark1999}, let $Y$ be a non-autonomous dynamical system
(the responder) evolving under the influence of an autonomous dynamical
system $X$ (the driver). $X$ is also called the driving or forcing system,
and $Y$ the driven or forced system. In the case of discrete-time
deterministic dynamical systems or flows observed at discrete times, this
situation is described by the difference equations%
\begin{equation}
\left\{ 
\begin{array}{l}
x_{t+1}=f(x_{t}) \\ 
y_{t+1}=g(x_{t},y_{t})%
\end{array}%
\right.  \label{dynamics}
\end{equation}%
where (\textit{i}) $x_{t}\in \mathcal{M}_{X}$ is the state of system $X$ at
time $t$, (\textit{ii}) $y_{t}\in \mathcal{M}_{Y}$ is the state of system $Y$
at time $t$, (\textit{iii}) $\mathcal{M}_{X}$ and $\mathcal{M}_{Y}$ are
compact manifolds of dimensions $\dim _{X}\geq 0$ and $\dim _{Y}\geq 1$,
respectively, (\textit{iv}) $f:\mathcal{M}_{X}\rightarrow \mathcal{M}_{X}$
is a $C^{1}$ diffeomorphism (i.e., a $C^{1}$ invertible map such that $%
f^{-1} $ is also $C^{1}$, where $C^{1}$ is shorthand for continuously
differentiable), and (\textit{v}) $g:\mathcal{M}_{X}\times \mathcal{M}%
_{Y}\rightarrow \mathcal{M}_{Y}$ is a $C^{1}$ map such that $g(x,\cdot )$ is
a diffeomorphism of $\mathcal{M}_{Y}$ for every $x\in \mathcal{M}_{X}$.
Alternatively, we say that there is a (unidirectional) coupling from $X$ to $%
Y$ and use the shorthand $X\rightsquigarrow Y$. Since we assume that $f$ and 
$g(x,\cdot )$ are invertible, we may take $t\in \mathbb{Z}$, although in
applications time series have a beginning that we will set at $t=0$.

\begin{remark}
By defining the map%
\begin{equation}
\lbrack f,g](x,y)=(f(x),g(x,y)),  \label{skew}
\end{equation}%
the forced system (\ref{dynamics}) becomes an autonomous dynamical system on
the full state space $\mathcal{M}_{X}\times \mathcal{M}_{Y}$, called the 
\emph{skew product} of $f$ and $g$. Due to the properties (ii)-(v) above, $%
[f,g]$ is a $C^{1}$ diffeomorphism of the compact manifold $\mathcal{M}%
_{X}\times \mathcal{M}_{Y}$.
\end{remark}

As stated in the Introduction, we are mainly interested in nonlinear time
series analysis. So, suppose further that the only information available
about the systems $X$ and $Y$ are scalar observations $\varphi _{X}(x_{t})$
of the states $x_{t}$ and $\varphi _{Y}(y_{t})$ of the states $y_{t}$, where
the observation functions $\varphi _{X}:\mathcal{M}_{X}\rightarrow \mathbb{R}
$ and $\varphi _{Y}:\mathcal{M}_{Y}\rightarrow \mathbb{R}$ are assumed to be 
$C^{1}$. To reconstruct the state spaces of the driver $X$ and the responder 
$Y$ from the corresponding observations $\varphi _{X}(x_{t})$ and $\varphi
_{Y}(y_{t})$, we use the Takens and Stark theorems respectively, which we
remind below for further reference.

\begin{theorem}
\label{ThmTakens1}\emph{[Takens Theorem \cite{Takens1981}]} If $d\geq 2\dim _{X}+1$%
, then the map $E_{f,\varphi _{X}}:\mathcal{M}_{X}\rightarrow \mathbb{R}^{d}$
defined as%
\begin{equation}
E_{f,\varphi _{X}}(x)=(\varphi _{X}(f^{0}(x)),\varphi
_{X}(f^{1}(x)),...,\varphi _{X}(f^{d-1}(x)))  \label{F}
\end{equation}%
is an embedding for generic $f$ and $\varphi _{X}$.
\end{theorem}

As usual, $f^{0}$ is the identity, $f^{1}=f$, and $f^{n}$ is the $n$-th
iterate of $f$. Here, \textquotedblleft generic $f$ and $\varphi _{X}$%
\textquotedblright\ formally means that the set $\{f,\varphi _{X}\}$ for
which $E_{f,\varphi _{X}}$ is an embedding (i.e., a $C^{1}$ diffeomorphism
onto its image) is open and dense in the $C^{1}$-topology (uniform
convergence of a map and its derivative) of the respective function spaces,
namely: diffeomorphisms of $\mathcal{M}_{X}$, and $C^{1}$ maps from $%
\mathcal{M}_{X}$ to $\mathbb{R}$. In general, a property is \textit{generic}
in a topological space $\mathcal{T}$ if it holds on a \textit{residual subset%
} $\mathcal{S}\subset \mathcal{T}$, i.e., on a subset that contains a
countable intersection of open sets. It turns out that an open and dense set
of maps $f$ for which $E_{f,\varphi _{X}}$ is an embedding for generic $%
\varphi _{X}$ is built by those $C^{1}$ diffeomorphisms of $\mathcal{M}_{X}$
that have only a finite number of periodic orbits of period less than $d$,
and the eigenvalues of each such periodic orbits are distinct (Stark \cite{Stark1999}, Theorem 2.2).

\begin{remark}
Theorem \ref{ThmTakens1} was generalized by Sauer et al. \cite{Sauer1991} in
two ways. First, by replacing \textquotedblleft generic\textquotedblright\
with \textquotedblleft probability-one\textquotedblright\ (in the sense of
prevalence). Second, by replacing the manifold $\mathcal{M}_{X}$ by a
compact invariant set $A$ that may have fractal box-counting dimension, and
the restriction $d\geq 2\dim _{X}+1$ (which comes from Whitney's Embedding
Theorem \cite{Whitney1936}) by $d\geq 2$\emph{boxdim}$(A)+1$, where \emph{%
boxdim}$(A)$ is the box-counting dimension of $A$.
\end{remark}

For our purposes, we need to generalize Theorem \ref{ThmTakens1} to the
forced dynamic 
\begin{equation*}
(x_{t+1},y_{t+1})=[f,g](x_{t},y_{t})
\end{equation*}%
defined by the diffeomorphism (\ref{skew}) in the full state space $\mathcal{%
M}_{X}\times \mathcal{M}_{Y}$. As before, set $[f,g]^{0}=$ identity, $%
[f,g]^{1}=[f,g]$ and $[f,g]^{t+1}=[f,g]\circ \lbrack f,g]^{t}$ for the
iterates of $[f,g]$, so that 
\begin{equation}
\lbrack f,g]^{t}(x,y)=(f^{t}(x),g^{(t)}(x,y)),  \label{[f,g]n}
\end{equation}%
where $g^{(t)}(x,y):\mathcal{M}_{X}\times \mathcal{M}_{Y}\rightarrow 
\mathcal{M}_{Y}$ is recursively defined by $g^{(0)}(x,y)=y$ and 
\begin{equation}
g^{(t)}(x,y)=g(f^{t-1}(x),g^{(t-1)}(x,y))  \label{g(n)(x,y)}
\end{equation}%
for $t\geq 1$. Application of the Takens Theorem to the skew product $[f,g]$
would provide a map $E_{[f,g],\varphi _{X,Y}}:\mathcal{M}_{X}\times \mathcal{%
M}_{Y}\rightarrow \mathbb{R}^{D}$, with $D\geq 2(\dim _{X}+\dim _{Y})+1$,
which would be an embedding for open dense sets of diffeomorphisms of $%
\mathcal{M}_{X}\times \mathcal{M}_{Y}$ and observation $C^{1}$ maps $\varphi
_{X,Y}(x,y):\mathcal{M}_{X}\times \mathcal{M}_{Y}\rightarrow \mathbb{R}$, in
the $C^{1}$ topology of the respective function spaces. However, what we
need for applications to nonlinear time series analysis is an embedding for
generic maps $f$, $g$ and observation maps $\varphi _{Y}$ on $\mathcal{M}%
_{Y} $, and this is not guaranteed by this approach.

The generalization of the Takens Theorem to forced dynamics that we need is
the following, due to Stark.

\begin{theorem}
\label{ThmForcedTakens}\emph{[Forced Takens Theorem \cite{Stark1999}]} If $D\geq
2(\dim _{X}+\dim _{Y})+1$, then the map $E_{f,g,\varphi _{Y}}:\mathcal{M}%
_{X}\times \mathcal{M}_{Y}\rightarrow \mathbb{R}^{D}$ defined as 
\begin{eqnarray}
&&E_{f,g,\varphi _{Y}}(x,y) \label{G}\\
&=&(\varphi _{Y}(g^{(0)}(x,y)),\,\varphi
_{Y}(g^{(1)}(x,y)),...,\,\varphi _{Y}(g^{(D-1)}(x,y))) \nonumber  
\end{eqnarray}
is an embedding for generic $f$, $g$ and $\varphi _{Y}$.
\end{theorem}

Specifically, generic $g$ means that $E_{f,g,\varphi _{Y}}$ is an embedding
for an open and dense set of diffeomorphisms $g(x,y)$ (such that $g(x,\cdot
) $ is a diffeomorphism of $\mathcal{M}_{Y}$ for every $x\in \mathcal{M}_{X}$%
) in the $C^{1}$-topology of $\mathcal{M}_{X}\times \mathcal{M}_{Y}$. In
this case, an open and dense set of maps $f$ for which $E_{f,g,\varphi _{Y}}$
is an embedding for generic $g$ and $\varphi _{Y}$ is built by those $C^{1}$
diffeomorphisms of $\mathcal{M}_{X}$ whose periodic orbits of period less
than $2d$ are isolated and have distinct eigenvalues (Stark \cite{Stark1999}, Theorem 3.1).


\subsection{The cross map}

\label{sec2.2}

Hereinafter we tacitly assume that $f$, $g$, $\varphi _{X}$ and $\varphi
_{Y} $ are generic in the sense of Theorems \ref{ThmTakens1} and \ref%
{ThmForcedTakens}. Also, \textquotedblleft smooth\textquotedblright\ stands
for $C^{1}$ smoothness in the following.

Given the scalar observations $(\varphi _{X}(x_{t}))_{t\in \mathbb{Z}}$ and $%
(\varphi _{Y}(y_{t}))_{t\in \mathbb{Z}}$, Theorems \ref{ThmTakens1} and \ref%
{ThmForcedTakens} allow to \textquotedblleft reconstruct\textquotedblright\
the (possibly unknown) dynamics of the underlying systems $X$ and $Y$ in the
manifolds%
\begin{equation*}
\mathcal{N}_{X}=E_{f,\varphi _{X}}(\mathcal{M}_{X})\subset \mathbb{R}^{d}
\end{equation*}%
and%
\begin{equation*}
\mathcal{N}_{Y}=E_{f,g,\varphi _{Y}}(\mathcal{M}_{X}\times \mathcal{M}%
_{Y})\subset \mathbb{R}^{D},
\end{equation*}%
called the \textit{reconstructed driver }and\textit{\ responder state spaces}%
, respectively, by means of the \textit{time delay vectors} 
\begin{equation}
\mathbf{x}_{t}=E_{f,\varphi _{X}}(x_{t})=(\varphi _{X}(x_{t}),\varphi
_{X}(x_{t+1}),...,\varphi _{X}(x_{t+d-1}))\in \mathbb{R}^{d}  \label{x_i}
\end{equation}%
and%
\begin{equation}
\mathbf{y}_{t}=E_{f,g,\varphi _{Y}}(x_{t},y_{t})=(\varphi
_{Y}(y_{t}),\varphi _{Y}(y_{t+1}),...,\varphi _{Y}(y_{t+D-1}))\in \mathbb{R}%
^{D}.  \label{y_i}
\end{equation}%
In turn, the dynamics $x_{t+1}=f(x_{t})$ in $\mathcal{M}_{X}$ translates
into the \textit{reconstructed driving dynamics} 
\begin{equation}
\mathbf{x}_{t+1}=(E_{f,\varphi _{X}}\circ f\circ E_{f,\varphi _{X}}^{-1})(%
\mathbf{x}_{t})=:\tilde{f}(\mathbf{x}_{t})  \label{f in N_X}
\end{equation}%
in $\mathcal{N}_{X}$, while the dynamics $%
(x_{t+1},y_{t+1})=[f,g](x_{t},y_{t})$ in $\mathcal{M}_{X}\times \mathcal{M}%
_{Y}$ translates into the \textit{reconstructed coupled dynamics} 
\begin{equation}
\mathbf{y}_{t+1}=(E_{f,g,\varphi _{Y}}\circ \lbrack f,g]\circ E_{f,g,\varphi
_{Y}}^{-1})(\mathbf{y}_{t})=:\widetilde{[f,g]}(\mathbf{y}_{t})
\label{[f,g] in N_X x N_Y}
\end{equation}%
in $\mathcal{N}_{Y}$, the manifolds $\mathcal{N}_{X}$ and $\mathcal{N}_{Y}$
being diffeomorphic copies of $\mathcal{M}_{X}$ and $\mathcal{M}_{X}\times 
\mathcal{M}_{Y}$, respectively. Therefore, all coordinate-independent
properties of $f$ and $[f,g]$ can be determined in $\mathcal{N}_{X}$ and $%
\mathcal{N}_{Y}$.

\begin{remark}
Without loss of generality, it can be assumed that $d=D$. In nonlinear time
series analysis, where the underlying dynamical system is unknown, the
embedding dimension of a time series is usually chosen by the method of
false nearest neighbors \cite{Kennel1992}.
\end{remark}

Let $\Pi _{X}:\mathcal{M}_{X}\times \mathcal{M}_{Y}\rightarrow \mathcal{M}%
_{X}$ be the projection onto $\mathcal{M}_{X}$, i.e., $\Pi _{X}(x,y)=x$.
From the diagram%
\begin{equation}
\begin{array}{rcccl}
\mathcal{M}_{X}\times \mathcal{M}_{Y}\ni & (x_{t},y_{t}) & \overset{\Pi _{X}}%
{\longrightarrow } & x_{t} & \in \mathcal{M}_{X} \\ 
E_{f,g,\varphi _{Y}}^{-1} & \uparrow &  & \downarrow & E_{f,\varphi _{X}} \\ 
\mathcal{N}_{Y}\ni & \mathbf{y}_{t} &  & \mathbf{x}_{t} & \in \mathcal{N}_{X}%
\end{array}
\label{chain}
\end{equation}%
along with the smoothness of the embeddings $E_{f,g,\varphi
_{Y}}^{-1},E_{f,\varphi _{X}}$ and the projection $\Pi _{X}$, we conclude
the following proposition.

\begin{proposition}
\label{PropositionCrossMap} A unidirectional coupling $X$ $\rightsquigarrow
Y $ necessarily implies the existence of a smooth map%
\begin{equation}
\Phi :=E_{f,\varphi _{X}}\circ \Pi _{X}\circ E_{f,g,\varphi _{Y}}^{-1}:%
\mathcal{N}_{Y}\rightarrow \mathcal{N}_{X}  \label{Phi}
\end{equation}%
called the \emph{cross map} of the coupling $X$ $\rightsquigarrow Y$, that
sends $\mathbf{y}_{t}$ to $\mathbf{x}_{t}$, i.e., 
\begin{equation}
\mathbf{x}_{t}=\Phi (\mathbf{y}_{t}).  \label{closeness map}
\end{equation}
\end{proposition}

Intuitively, equation (\ref{closeness map}) spells out that the responder
signal carries information about the dynamics of the driver because of the
time evolution law $y_{t+1}=g(x_{t},y_{t})$.

\begin{remark}
\label{Remark Phi(k)}Equation (\ref{closeness map}) is equivalent to the
existence of a map 
\begin{equation}
\mathbf{x}_{t}=\Phi ^{(k)}(\mathbf{y}_{t-k}),  \label{Cross(k) 1}
\end{equation}%
for any $k\in \mathbb{Z}$, where $\Phi ^{(k)}:\mathcal{N}_{Y}\rightarrow 
\mathcal{N}_{X}$. Indeed, from 
\begin{equation}
\mathbf{x}_{t}=\tilde{f}^{k}(\mathbf{x}_{t-k})  \label{Cross(k) 2}
\end{equation}%
(see equation (\ref{f in N_X})) and $\mathbf{x}_{t-k}=\Phi (\mathbf{y}%
_{t-k}) $, it follows 
\begin{equation}
\Phi ^{(k)}=\tilde{f}^{k}\circ \Phi =E_{f,\varphi _{X}}\circ f^{k}\circ
E_{f,\varphi _{X}}^{-1}\circ \Phi  \label{Cross(k) 3}
\end{equation}%
and, hence, 
\begin{equation}
\mathbf{x}_{t}=\frac{1}{K}\sum_{k=0}^{K-1}\Phi ^{(k)}(\mathbf{y}%
_{t-k})=:\Phi _{K}(\mathbf{y}_{t},\mathbf{y}_{t-1},...,\mathbf{y}_{t-K+1})
\label{Cross(k) 4}
\end{equation}%
for all $K\geq 1$. Note that $\Phi ^{(k)}$ and $\Phi _{K}$ are continuous,
and $\Phi ^{(0)}=\Phi _{1}=\Phi $.
\end{remark}

By changing the summation limits in (\ref{Cross(k) 4}), one can construct
other similar multi-time expressions. For definiteness, we will use only the
definition (\ref{Cross(k) 4}).

\begin{definition}
We call the continuous map $\mathbf{x}_{t}=\Phi ^{(k)}(\mathbf{y}_{t-k})$
the \emph{cross map of order} $k\in \mathbb{Z}$, and the continuous map $%
\mathbf{x}_{t}=\Phi _{K}(\mathbf{y}_{t},\mathbf{y}_{t-1},...,\mathbf{y}%
_{t-K+1})$ the \emph{cross map of period }$K\geq 1$.
\end{definition}

The continuity of the cross map $\Phi $ has been used in nonlinear time
series analysis to discriminate functional (deterministic, causal, ...)
relationships between observations due to coupled dynamics from statistical
correlation. In its simplest version, the continuity of the cross map $%
\mathbf{x}_{t}=\Phi (\mathbf{y}_{t})$ belonging to the coupling $%
X\rightsquigarrow Y$ implies that, given an open ball $B_{\varepsilon }(%
\mathbf{x}_{t})\subset \mathbb{R}^{d}$ with center $\mathbf{x}_{t}$ and
arbitrary radius $\varepsilon >0$, there exists an open ball $B_{\delta }(%
\mathbf{y}_{t})\subset \mathbb{R}^{D}$ with center $\mathbf{y}_{t}$ and
radius $\delta =\delta (\varepsilon )>0$ such that $\Phi (B_{\delta }(%
\mathbf{y}_{t}))\subset B_{\varepsilon }(\mathbf{x}_{t})$. Therefore, the $k$
nearest neighbors $\mathbf{y}_{t_{1}},...,\mathbf{y}_{t_{k}}$ of a time
delay vector $\mathbf{y}_{t}\in \mathcal{N}_{Y}$ in a time series $(\mathbf{y%
}_{t})_{0\leq t\leq T}$ of the responder are mapped by $\Phi $ to close
neighbors $\mathbf{x}_{t_{1}},...,\mathbf{x}_{t_{k}}$ of the contemporaneous
vector $\mathbf{x}_{t}\in \mathcal{N}_{Y}$ in the corresponding time series $%
(\mathbf{x}_{t})_{0\leq t\leq T}$ of the driver. Methods that take advantage
of the continuity of $\Phi $ in this way to unveil the coupling $%
X\rightsquigarrow Y$ include \textit{cross prediction} \cite{LeVanQuyen1999}%
, \textit{convergent cross mapping} \cite{Sugihara2012} and \textit{%
continuity scaling} \cite{Ying2022}.


\subsection{Generalized synchronization}

\label{sec2.3}

According to Rulkov et al. \cite{Rulkov1995} and Pikovsky et al. \cite{Pikovsky2001}, the systems $X\rightsquigarrow
Y $ are in \textit{generalized }(or \textit{general}) \textit{synchronization%
} if there exists a continuous map $h:$ $\mathcal{M}_{X}\rightarrow \mathcal{%
M}_{Y}$ such that 
\begin{equation}
y_{t}=h(x_{t})  \label{sync map}
\end{equation}%
for all $t\in \mathbb{Z}$. That is, the responder follows the driver but in
a weaker form than in identical synchronization, which corresponds to $h$
being the identity (i.e., $X$ and $Y$ are structurally the same and $%
y_{t}=x_{t}$). We will also say that $Y$ is synchronized to $X$ if equation (%
\ref{sync map}) holds and call $h(x)$ the \textit{synchronization map}.

Therefore, in case of synchronization the full state space $\mathcal{M}%
_{X}\times \mathcal{M}_{Y}$ shrinks into the subspace $\{(x,y)\in \mathcal{M}%
_{X}\times \mathcal{M}_{Y}:y=h(x)\}$, which is the graph of the
synchronization map $x\mapsto h(x)$. This subspace is usually called the 
\textit{synchronization} \textit{manifold}, even when $h$ is not smooth. It
follows then that the projection map from $\mathcal{M}_{X}\times \mathcal{M}%
_{Y}$ onto $\mathcal{M}_{X}$, $\Pi _{X}(x,y)=x$, is invertible, 
\begin{equation}
\Pi _{X}^{-1}(x)=(x,h(x)),  \label{Pi(-1) = h}
\end{equation}%
and the range of $\Pi _{X}^{-1}:\mathcal{M}_{X}\rightarrow \mathcal{M}%
_{X}\times \mathcal{M}_{Y}$ is the synchronization manifold.

\begin{remark}
\label{Remark h(k)}Plug the driver dynamic $x_{t}=f(x_{t-1})$ into equation (%
\ref{sync map}) to derive 
\begin{equation}
y_{t}=h(x_{t})=(h\circ f)(x_{t-1})=...=(h\circ
f^{k})(x_{t-k})=:h^{(k)}(x_{t-k}),  \label{sync map B}
\end{equation}%
where $h^{(k)}=h\circ f^{k}:$ $\mathcal{M}_{X}\rightarrow \mathcal{M}_{Y}$
is continuous for all $k\geq 0$, and $h^{(0)}=h$. Equation (\ref{sync map B}%
) with $k\geq 1$ corresponds to generalized synchronization for responders
with an internal delay loop. When $f$ is invertible (as in our case), the
generalized synchronization of the systems $X$ and $Y$ can indistinctly be
defined by $y_{t}=h(x_{t})$ or, more generally, by $y_{t}=h^{(k)}(x_{t-k})$
for $k\in \mathbb{Z}$; in the latter case, $h=h^{(k)}\circ f^{-k}$. The maps 
$h^{(k)}:$ $\mathcal{M}_{X}\rightarrow \mathcal{M}_{Y}$ are called \emph{%
synchronization maps of order} $k$, orders $0$ and $1$ being the usual
choices in applications.
\end{remark}

From equation (\ref{sync map B}) it trivially follows that%
\begin{equation}
y_{t}=\frac{1}{K}%
\sum_{k=0}^{K-1}h^{(k)}(x_{t-k})=:h_{K}(x_{t},x_{t-1},...,x_{t-K+1})
\label{multiv h}
\end{equation}%
for all $K\geq 1$, in case $Y$ is synchronized with $X$. By changing the
summation limits in equation (\ref{multiv h}), one can construct other
similar expressions. For definiteness, we will use only equation (\ref%
{multiv h}) in this paper, so that $h_{1}=h$.

Therefore, to detect synchronization of a time series $\{y_{t}\}_{t\geq 0}$
with another time series $\{x_{t}\}_{t\geq 0}$, we can look for functional
dependencies of the form (\ref{multiv h}) with $K>1$ rather than $%
y_{t}=h(x_{t})$. If $\{x_{t}\}_{t\geq 0}$ is a deterministic time series
(i.e., $x_{t+1}=f(x_{t})$) and $\{y_{t}\}_{t\geq 0}$ is synchronized with
it, then equation (\ref{multiv h}) holds with a continuous map $h_{K}:%
\mathcal{M}_{X}^{K}\rightarrow \mathcal{M}_{Y}$ such that $%
h_{K}(x_{t},f^{-1}(x_{t}),...,f^{-K+1}(x_{t}))=h(x_{t})$. The point is that,
in time series analysis, multi-time dependencies like (\ref{multiv h}) can
be efficiently detected by recurrent neural nets, as we discuss in Section %
\ref{sec6}.

\begin{definition}
\label{DefMultivariateSyncMap}We call the continuous map $%
y_{t}=h_{K}(x_{t},...,x_{t-K+1})$ in equation (\ref{multiv h}) the \emph{%
synchronization map of period} $K\geq 1$.
\end{definition}

The synchronization maps of order $k$, $h^{(k)}=h\circ f^{k}$, satisfy a
number of straightforward relations involving also the function $g(x,y)$.
Indeed, in case of synchronization, the dynamic (\ref{dynamics}) of $%
X\rightsquigarrow Y$ simplifies to%
\begin{equation}
\left\{ 
\begin{array}{l}
x_{t+1}=f(x_{t}) \\ 
y_{t+1}=g(x_{t},h(x_{t}))%
\end{array}%
\right.  \label{dynamics sync}
\end{equation}%
Comparing with $y_{t+1}=h(f(x_{t}))$ shows that $h(x)$ fulfills the
functional relation%
\begin{equation}
h(f(x))=g(x,h(x)).  \label{sync map E2}
\end{equation}%
Replace $x$ with $f^{k}(x)$ in equation (\ref{sync map E2}) to obtain%
\begin{equation}
h^{(k+1)}(x)=h^{(k)}(f(x))=g(f^{k}(x),h^{(k)}(x)).  \label{sync map F}
\end{equation}%
Recursion of equation (\ref{sync map F}) leads to alternative formulas for
synchronization maps of arbitrary periods involving the function $g$.

Contingent upon the structure of $g(x,y)$, the synchronization map $h(x)$
can sometimes be written in closed form; see Pikovsky et al. \cite{Pikovsky2001} and Parlitz \cite{Parlitz2012}
for an example with a baker map. Interestingly, the parameters of
that example can be fine tuned so that the cross sections $x^{(2)}=\mathrm{%
const}$ of $h(x^{(1)},x^{(2)})$ are Weierstrass functions, i.e., continuous
functions that are nowhere differentiable.

The definition of synchronization can be weakened by requiring the condition
(\ref{sync map}) only asymptotically. In more formal terms, we say that the
responder $Y $ is \textit{asymptotically synchronized} to the driver $X$ if
there exists a continuous map $h:$ $\mathcal{M}_{X}\rightarrow \mathcal{M}%
_{Y}$ such that%
\begin{equation}
\lim_{t\rightarrow \infty }\left\Vert y_{t}-h(x_{t})\right\Vert =0,
\label{weak sync}
\end{equation}%
where $\left\Vert \cdot \right\Vert $ is a distance in $\mathcal{M}_{Y}$. In
this case, the synchronization manifold becomes an attracting set in $%
\mathcal{M}_{X}\times \mathcal{M}_{Y}$.

A direct consequence of asymptotic synchronization is the \textit{asymptotic
stability} of the responder. We say that the responder $Y$ is asymptotically
stable if all orbits converge to the same orbit regardless of the initial
condition, that is, if given two responses $(y_{t})_{t\geq 0}$ and $(\tilde{y%
}_{t})_{t\geq 0}$ to a signal $(x_{t})_{t\geq 0}$ from the driver with
different initial conditions $y_{0}\neq \tilde{y}_{0}$, then%
\begin{equation}
\lim_{t\rightarrow \infty }\left\Vert y_{t}-\tilde{y}_{t}\right\Vert =0.
\label{aux syst}
\end{equation}

Asymptotic stability of the responder is weaker than asymptotic
synchronization because the existence of a hypothetical synchronization map
does not follow from equation (\ref{aux syst}). This is the case, for
example, when a periodic driver has gone through a period-doubling
bifurcation \cite{Parlitz1997} or there is a multistability in the
responder, i.e, a driver signal $(x_{t})_{t\geq 0}$ can elicit two or more
stable responses \cite{Pikovsky2001}.

On the other hand, the asymptotic stability of the responder provides a
simple method to test synchronization called the \textit{auxiliary system
method} \cite{Abarbanel1996}. This method boils down to check equation (\ref%
{aux syst}) for two initial conditions $y_{0}\neq \tilde{y}_{0}$; if (\ref%
{aux syst}) does not hold, then $Y$ is not synchronized to $X$.


\subsection{Relationship between generalized synchronization and the cross
map}

\label{sec2.4}

According to Proposition \ref{PropositionCrossMap}, a coupling $%
X\rightsquigarrow Y$ implies the existence of the cross map $\mathbf{x}%
_{t}=\Phi (\mathbf{y}_{t})$, whereas the synchronization map $y_{t}=h(x_{t})$
exists in seemingly exceptional cases (unless the coupling is strong
enough). Despite this notable difference, both maps are closely related, as
we will now see.

Let $\Pi _{Y}(x,y)=y$ be the projection map from $\mathcal{M}_{X}\times 
\mathcal{M}_{Y}$ onto $\mathcal{M}_{Y}$. On the one hand, from the diagram%
\begin{equation}
\begin{array}{rcccl}
\mathcal{M}_{X}\ni & x_{t} & \overset{h}{\longrightarrow } & y_{t} & \in 
\mathcal{M}_{Y} \\ 
E_{f,\varphi _{X}} & \downarrow &  & \uparrow & \Pi _{Y}\circ E_{f,g,\varphi
_{Y}}^{-1} \\ 
\mathcal{N}_{X}\ni & \mathbf{x}_{t} & \overset{\Phi }{\longleftarrow } & 
\mathbf{y}_{t} & \in \mathcal{N}_{Y}%
\end{array}
\label{diagram}
\end{equation}%
we have that, if $\Phi $ is invertible, then $h$ exists and%
\begin{equation}
h=\Pi _{Y}\circ E_{f,g,\varphi _{Y}}^{-1}\circ \Phi ^{-1}\circ E_{f,\varphi
_{X}}.  \label{h Phi}
\end{equation}

On the other hand, equation (\ref{Pi(-1) = h}) spells out that, if $h$
exists, then $\Pi _{X}$ is invertible. From equation (\ref{Phi}) it follows
then that%
\begin{equation}
\Phi ^{-1}=E_{f,g,\varphi _{Y}}\circ \Pi _{X}^{-1}\circ E_{f,\varphi
_{X}}^{-1}.  \label{Phi-1}
\end{equation}%
The bottom line of equations (\ref{h Phi}) and (\ref{Phi-1}) is the
following.

\begin{proposition}
\label{ThmCrossSyncMaps}The systems $X\rightsquigarrow Y$ are synchronized
if and only if the cross map $\mathbf{x}_{t}=\Phi (\mathbf{y}_{t})$ is
invertible and bicontinuous (i.e., $\Phi ^{-1}$ is continuous). In this
case, the synchronization map $y_{t}=h(x_{t})$ and $\mathbf{y}_{t}=\Phi
^{-1}(\mathbf{x}_{t})$ are related through the expressions (\ref{h Phi}) and
(\ref{Phi-1}), where $\Pi _{X}^{-1}(x)=(x,h(x))$.
\end{proposition}

In other words, $\Phi ^{-1}:\mathcal{N}_{X}\rightarrow \mathcal{N}_{Y}$ is
the synchronization map of the systems $X\rightsquigarrow Y$ in the
reconstructed state spaces (if it exists and is continuous); see Pecora et al. \cite%
{Pecora1995} for numerical methods to test whether two time series are
related by a map with properties such as continuity, invertibility,
smoothness and more. As a rule, the relationship $x\mapsto (x,y)$ is
multivalued owing to folds in the manifold $\mathcal{M}_{X}\times \mathcal{M}%
_{Y}$, so generalized synchronization is rather an exception. Multivalued
synchronization maps, corresponding to noninvertible cross maps, have been
considered, e.g., in Rulkov et al. \cite{Rulkov2001} and Parlitz \cite{Parlitz2012}.

To lift $y_{t}=h_{K}(x_{t},x_{t-1},...,x_{t-K+1})$, the synchronization map
of period $K$ (\ref{multiv h}), to the reconstructed state spaces $\mathcal{N%
}_{X}$ and $\mathcal{N}_{Y}$, use the reconstructed driver dynamic $\tilde{f}%
:\mathcal{N}_{X}\rightarrow \mathcal{N}_{X}$ defined in equation (\ref{f in
N_X}) to derive%
\begin{equation}
\mathbf{y}_{t}=\Phi ^{-1}(\mathbf{x}_{t})=(\Phi ^{-1}\circ \tilde{f})(%
\mathbf{x}_{t-1})=...=(\Phi ^{-1}\circ \tilde{f}^{k})(\mathbf{x}_{t-k}),
\label{multi h reconst 1B}
\end{equation}%
where $k\in \mathbb{Z}$ and 
\begin{equation}
\Phi ^{-1}\circ \tilde{f}^{k}=\Phi ^{-1}\circ E_{f,\varphi _{X}}\circ
f^{k}\circ E_{f,\varphi _{X}}^{-1}=(\Phi ^{(-k)})^{-1},
\label{multi h reconst 2B}
\end{equation}%
by the definition of the cross map of order $k$, equation (\ref{Cross(k) 3}%
). Therefore, the synchronization map of period $K\geq 1$ (\ref{multiv h})
becomes 
\begin{equation}
\mathbf{y}_{t}=\frac{1}{K}\sum_{k=0}^{K-1}(\Phi ^{(-k)})^{-1}(\mathbf{x}%
_{t-k})=:H_{K}(\mathbf{x}_{t},\mathbf{x}_{t-1},...,\mathbf{x}_{t-K+1})
\label{multi h reconst 2}
\end{equation}%
in the reconstructed spaces. Note that $H_{1}=\Phi ^{-1}$.

\begin{definition}
The continuous map $\mathbf{y}_{t}=H_{K}(\mathbf{x}_{t},\mathbf{x}_{t-1},...,%
\mathbf{x}_{t-K+1})$ defined in equation (\ref{multi h reconst 2}) will be
called the \emph{reconstructed synchronization map of period} $K\geq 1.$
\end{definition}

We will harness $H_{K}$ with $K>1$ in the applications with synthetic data
(Section \ref{sec6}) and real world data (Section \ref{sec7}) via recurrent
neural networks. We remark already at this point that, unlike the synthetic
data of Section \ref{sec6}, real world data are generally bidirectionally
coupled, as happens with the EEGs of Section \ref{sec7}. Following the
standard approach, we will measure the coupling strength between pairs of
EEGs in both directions, the \textquotedblleft driver\textquotedblright\
being identified by the direction with the strongest coupling. In case of
equal strengths, the systems are assumed to be synchronized.

As mentioned in Section \ref{sec2.2}, in case of unidirectional coupling
(the framework of this paper) the relationship $\mathbf{x}_{t}=\Phi (\mathbf{%
y}_{t})$ due to the coupled dynamic $X\rightsquigarrow Y$ can be unveiled
numerically from the time series $(\mathbf{x}_{t})_{0\leq t\leq T}$ and $(%
\mathbf{y}_{t})_{0\leq t\leq T}$ \cite{Pecora1995,Sugihara2012,Ying2022}.
Thus, in the method of nearest-neighbor cross prediction, one estimates $%
\mathbf{x}_{t}$ or $\mathbf{x}_{t+1}$ based on the nearest neighbors of $%
\mathbf{y}_{t}$ for any $0\leq t\leq T$ to test for the existence of $\Phi $%
. Likewise, if $\Phi ^{-1}$ exists and is continuous (i.e., $X$ and $Y$ are
synchronized), then one can also discern the inverse relationship $\mathbf{y}%
_{t}=\Phi ^{-1}(\mathbf{x}_{t})$ by the same techniques. As a matter of
fact, in the case of a bijective and bicontinuous $\Phi $, there is a 1-to-1
relation between the neighborhoods of nearest neighbors of $\mathbf{y}_{t}$
and $\mathbf{x}_{t}$, so, if $Q_{T}(\Phi )$ and $Q_{T}(\Phi ^{-1})$ are
fidelity metrics of the respective estimations, then 
\begin{equation}
Q_{T}(\Phi )-Q_{T}(\Phi ^{-1})\simeq 0,  \label{CT - CT}
\end{equation}%
where it applies that the longer the time series, the better the predictions
and, hence, the smaller $Q_{T}(\Phi )-Q_{T}(\Phi ^{-1})$. In other words,
the continuity of the cross map and its inverse can be exploited via (\ref%
{CT - CT}) to test two time series $(\mathbf{x}_{t})_{0\leq t\leq T}$ and $(%
\mathbf{y}_{t})_{0\leq t\leq T}$ for general synchronization.

\begin{remark}
The nonexistence of the cross map can uncover common drivers. Indeed, if $%
Z\rightsquigarrow X$ and $Z\rightsquigarrow Y$ then $\mathbf{z}_{t}=\Phi (%
\mathbf{x}_{t})$ and $\mathbf{z}_{t}=\tilde{\Phi}(\mathbf{y}_{t})$, so that $%
\Phi (\mathbf{x}_{t})=\tilde{\Phi}(\mathbf{y}_{t})$. Here $\Phi :\mathcal{N}%
_{X}\rightarrow \mathcal{N}_{Z}$ and $\tilde{\Phi}:\mathcal{N}%
_{Y}\rightarrow \mathcal{N}_{Z}$ are the cross maps associated to the
coupled dynamics $Z\rightsquigarrow X$ and $Z\rightsquigarrow Y$,
respectively. But there is no cross map between $\mathcal{N}_{X}$ and $%
\mathcal{N}_{Y}$, unless $\Phi $ or $\tilde{\Phi}$ is invertible and
continuous (so that $\mathbf{x}_{t}=(\Phi ^{-1}\circ \tilde{\Phi})(\mathbf{y}%
_{t})$ or $\mathbf{y}_{t}=(\tilde{\Phi}^{-1}\circ \Phi )(\mathbf{x}_{t})$),
in which case $X$ or $Y$ is synchronized with $Z$.
\end{remark}

To wrap up this section, let us point out that the diagram (\ref{diagram})
is a particularization of the diagram 
\begin{equation}
\begin{array}{rcccl}
\mathcal{M}_{X}\ni & x_{t} & \overset{h^{(k)}}{\longrightarrow } & y_{t+k} & 
\in \mathcal{M}_{Y} \\ 
E_{f,\varphi _{X}} & \downarrow &  & \uparrow & \Pi _{Y}\circ E_{f,g,\varphi
_{Y}}^{-1} \\ 
\mathcal{N}_{X}\ni & \mathbf{x}_{t} & \overset{\Phi ^{(-k)}}{\longleftarrow }
& \mathbf{y}_{t+k} & \in \mathcal{N}_{Y}%
\end{array}
\label{diagram2}
\end{equation}%
to $k=0$, where $h^{(k)}$ is the synchronization map of order $k$ (equation (%
\ref{sync map B}), $h^{(0)}=h$) and $\Phi ^{(-k)}$ is the cross map of order 
$-k$ (equation (\ref{Cross(k) 1}), $\Phi ^{(0)}=\Phi $). Thus, equation (\ref%
{h Phi}) is the special case $k=0$ of the relationship%
\begin{eqnarray}
h^{(k)} &=&\Pi _{Y}\circ E_{f,g,\varphi _{Y}}^{-1}\circ (\Phi
^{(-k)})^{-1}\circ E_{f,\varphi _{X}}  \label{h(k) Phi(-1)} \\
&=&\Pi _{Y}\circ E_{f,g,\varphi _{Y}}^{-1}\circ \Phi ^{-1}\circ E_{f,\varphi
_{X}}\circ f^{k}  \notag
\end{eqnarray}%
where we used (\ref{multi h reconst 2B}) and (\ref{f in N_X}) in the second
line. In this regard, note that $h^{(k)}$ is invertible if and only if $h$
is invertible (since $h^{(k)}=h\circ f^{k}$ by equation (\ref{sync map B})$)$
and, likewise, $\Phi ^{(-k)}$ is invertible if and only if $\Phi $ is
invertible (since $\Phi ^{(-k)}=E_{f,\varphi _{X}}\circ f^{-k}\circ
E_{f,\varphi _{X}}^{-1}\circ \Phi $ by equation (\ref{Cross(k) 3})).


\section{Dynamical noise as stochastic forcing}

\label{sec3}

Random or \textquotedblleft noisy\textquotedblright\ dynamical systems can
be modeled in different ways, from the perhaps simplest ones, such as
switching systems \cite{Amigo2013B,Amigo2013C} and iterated function systems 
\cite{Barnsley1985} to nonautonomous dynamical systems \cite{Kloeden2011}
and full-fledged random dynamical systems, described by random differential
equations \cite{Arnold2003}. In our setting, a natural way to turn a
noiseless dynamic, say, $x_{t+1}=f(x_{t})$ on a compact manifold $\mathcal{M}%
_{X}$, into a noisy one is to replace the map $f$ with a family of maps $%
\{f_{\omega _{t}}\}_{t\in \mathbb{Z}}$, where the index $\omega _{t}$ is a
(possibly multi-component) parameter belonging to a suitable space that is
randomly chosen at each discrete time $t$. For example, $f_{\omega
_{t}}(x_{t})=f(x_{t})+\omega _{t}$, $\omega _{t}\in \mathcal{M}_{X}$, models
additive dynamical noise, while $f_{\omega _{t}}(x_{t})=\omega _{t}f(x_{t})$
models multiplicative dynamical noise. This approach, sometimes called
\textquotedblleft finitely parameterized stochasticity\textquotedblright\ 
\cite{Muldoon1998}, is sufficient for most practical applications \cite%
{Hirata2021}. So, the term \textit{noisy dynamical system} will refer
hereafter to such implementation of dynamical noise via stochastic processes
in the parameter space; our parameter spaces will be compact topological
sets.

At this point we recall that each \textit{stationary} (discrete-time)
stochastic process corresponds in a canonical way to a so-called \textit{%
shift system}, which is a dynamical system whose states are the realizations
of the stochastic process considered \cite{Petersen2000}. In other words,
stationary stochastic processes can be modeled as autonomous dynamical
systems. It is therefore not surprising that shift systems allow to
formulate noisy dynamical systems as forced systems in a manner formally
similar to the noiseless case. But before getting to that point, we need to
introduce the concepts and notation.

Let $\Omega $ be a compact topological space of parameters and let $\Omega ^{%
\mathbb{Z}}$ be the set of all two-sided sequences of elements of $\Omega $, 
\begin{equation*}
\omega =(\ldots ,\omega _{-t},\ldots ,\omega _{-1},\omega _{0},\omega
_{1},\ldots ,\omega _{t},\ldots ),
\end{equation*}%
endowed with the product topology. As a result, $\Omega ^{\mathbb{Z}}$ is a
compact topological space, too. Furthermore, let $\sigma :\Omega ^{\mathbb{Z}%
}\rightarrow \Omega ^{\mathbb{Z}}$ be the (left) shift map,%
\begin{equation*}
\sigma (\omega )=(\ldots ,\omega _{-t+1},\ldots ,\omega _{0},\overset{\ast }{%
\omega _{1}},\omega _{2},\ldots ,\omega _{t+1},\ldots ),
\end{equation*}%
where the asterisk marks the zeroth component; component-wise, $[\sigma
(\omega )]_{t}=\omega _{t+1}$ for all $t\in \mathbb{Z}$. The shift map is a
homeomorphism of $\Omega ^{\mathbb{Z}}$.

In addition, the continuous (or topological) dynamical system $(\Omega ^{%
\mathbb{Z}},\sigma )$ can be further promoted to a measure-preserving
dynamical system by introducing a $\sigma $-invariant (probability) measure $%
\mu $ on the Borel sigma-algebra $\mathcal{B}^{\times }$ of $\Omega ^{%
\mathbb{Z}}$ via the finite-dimensional probability distributions of the
given or desired $\Omega $-valued stochastic process \cite{Petersen2000}.
For instance, the $n$-dimensional marginal probabilities 
$\mathbb{P}(\omega_{i_{1}}\in B_{i_{1}},\ldots ,\omega_{i_{n}}\in
B_{i_{n}})$ on $(\Omega ^{%
\mathbb{Z}},\mathcal{B}^{\times })$ are given by 
\begin{equation}
\mu _{n}(B_{i_{1}}\times \ldots \times B_{i_{n}}):=\mu (B_{i_{1}}\times
\ldots \times B_{i_{n}}\times \underset{i\neq i_{1},...,i_{n}}{\prod }\Omega
) \label{mu()=Pr()}
\end{equation}%
where $B_{i_{1}},...,B_{i_{n}}$ are Borel sets (e.g., open sets) of $\Omega $%
. The resulting dynamical system $\Sigma =(\Omega ^{\mathbb{Z}},\mathcal{B}%
^{\times },\mu ,\sigma )$ is the \textit{shift system} mentioned above. When
the measure $\mu $ consists of finitely many atoms, then $\{f_{\omega
_{t}}\} $ is an iterated function system \cite{Barnsley1985}. Product
measures, 
\begin{equation*}
\mu _{n}(B_{i_{1}}\times B_{i_{2}}\times \ldots \times B_{i_{n}})=\mu
_{1}(B_{i_{1}})\mu _{1}(B_{i_{2}})\ldots \mu _{1}(B_{i_{n}}),
\end{equation*}%
correspond to independent (memoryless) processes such as coin tossing and
white noise.

Following Stark et al. \cite{Stark2003}, a noisy dynamical system $X$ is
then modeled by the skew product%
\begin{equation}
\left\{ 
\begin{array}{l}
\omega _{t+1}=[\sigma (\omega )]_{t} \\ 
x_{t+1}=f(\omega _{t},x_{t})=f_{\omega _{t}}(x_{t})%
\end{array}%
\right.  \label{noisy}
\end{equation}%
where we suppose that $f_{\omega _{t}}=f(\omega _{t},\cdot ):\mathcal{M}%
_{X}\rightarrow \mathcal{M}_{X}$ is a diffeomorphism for all $\omega _{t}\in
\Omega $. Alternatively,%
\begin{equation}
\left\{ 
\begin{array}{l}
\omega _{t+1}=[\sigma ^{t}(\omega )]_{0} \\ 
x_{t+1}=f([\sigma ^{t}(\omega )]_{0},x_{t})=f_{[\sigma ^{t}(\omega
)]_{0}}(x_{t})%
\end{array}%
\right.  \label{noisy2}
\end{equation}%
i.e., the parameter of the dynamic at time $t$ is the $0$-component of the
shifted sequence $\sigma ^{t}(\omega )$.

Due to the formal similarity of equation (\ref{noisy}) with equation (\ref%
{dynamics}) for the forced dynamic $X\rightsquigarrow Y$, the modeling (\ref%
{noisy}) of a noisy dynamical system is called \textit{stochastic forcing} 
\cite{Stark2003}. Indeed, here we have $\Sigma \rightsquigarrow X$, where
the shift system $\Sigma =(\Omega ^{\mathbb{Z}},\mathcal{B}^{\times },\mu
,\sigma )$ is also an autonomous dynamical system and $X$ is \textit{randomly%
} forced by $\Sigma $ since $x_{t+1}=f(\omega _{t},x_{t})$. This parallelism
also carries over to the embedding $E_{f,\varphi _{X}}:\mathcal{M}%
_{X}\rightarrow \mathbb{R}^{d}$, equation (\ref{F}), as follows.

Let $\omega =(\omega _{t})_{t\in \mathbb{Z}}$ be a two-sided sequence of
points in $\Omega $ and set%
\begin{equation}
f_{\omega _{t},...,\omega _{0}}(x)=(f_{\omega _{t}}\circ ...\circ f_{\omega
_{0}})(x),  \label{f_omegas}
\end{equation}%
for all $x\in \mathcal{M}_{X}$, so $f_{\omega _{t},...,\omega _{0}}:\mathcal{%
M}_{X}\rightarrow \mathcal{M}_{X}$ for all $t\geq 0$. For every $\omega $,
define the map $E_{f,\varphi _{X},\omega }:\mathcal{M}_{X}\rightarrow 
\mathbb{R}^{d}$ as%
\begin{equation}
E_{f,\varphi _{X},\omega }(x)=(\varphi _{X}(x),\,\varphi _{X}(f_{\omega
_{0}}(x)),...,\,\varphi _{X}(f_{\omega _{d-2},..,\omega _{0}}(x))).
\label{F omega}
\end{equation}%
Note that $E_{f,\varphi _{X},\omega }$ actually depends on the $d-1$
parameters $\omega _{0},\omega _{1},...,\omega _{d-2}$.

\begin{theorem}
\label{ThmStarkOmega}\emph{[Stark et al. \cite{Stark2003}]} If $d\geq 2\dim _{X}+1$, then there
exists a residual set of $(f,\varphi _{X})$ such that for any $(f,\varphi
_{X})$ in this set there is an open dense set of sequences $\omega \in
\Omega ^{\mathbb{Z}}$ such that the map $E_{f,\varphi _{X},\omega }$ is an
embedding.
\end{theorem}

Finally, let us point out that Theorem \ref{ThmStarkOmega} generalizes
readily to the case of noisy observations. For example, if the observation
function $\varphi _{X}(x)$ is replaced by the noisy\ observation function $%
\varphi _{X,\eta }(x)=\varphi _{X}(x,\eta )$, where $\eta \in (\Omega
^{\prime })^{\mathbb{Z}}$, $\Omega ^{\prime }$ is a compact set and $\Sigma
^{\prime }$ is the corresponding shift space, then the map%
\begin{eqnarray}
&&E_{f,\varphi _{X},\omega ,\eta }(x) \label{F omega eta} \\
&=&(\varphi _{X,\eta _{0}}(x),\,\varphi
_{X,\eta _{1}}(f_{\omega _{0}}(x)),...,\,\varphi _{X,\eta _{d-1}}(f_{\omega
_{d-2},..,\omega _{0}}(x)))  \nonumber
\end{eqnarray}%
is an embedding for generic $(\omega ,\eta )\in \Sigma \times \Sigma
^{\prime }$. See Stark et al. \cite{Stark2003} for more detail and other possibilities.
Therefore, we may assume hereafter that the observations are noiseless for
notational simplicity.


\section{Coupled dynamics and noise}

\label{sec4}

Next we show that the skew product (\ref{noisy}) includes the case of two
unidirectionally coupled systems $X\rightsquigarrow Y$, where the driver is
a noisy dynamical system, namely,

\begin{equation}
\left\{ 
\begin{array}{l}
x_{t+1}=f_{\omega _{t}}(x_{t}) \\ 
y_{t+1}=g(x_{t},y_{t})%
\end{array}%
\right.  \label{forcing with dyn noise}
\end{equation}%
Equivalently,%
\begin{equation}
x_{t+1}=f_{\omega _{t},\omega _{t-1},...,\omega _{0}}(x_{0}),
\label{forcing with dyn noise 1}
\end{equation}%
see equation (\ref{f_omegas}), and%
\begin{equation}
y_{t+1}=(g_{x_{t}}\circ g_{x_{t-1}}\circ ...\circ
g_{x_{0}})(y_{0})=g_{x_{t},x_{t-1},...,x_{0}}(y_{0})
\label{forcing with dyn noise 2}
\end{equation}%
for $t\geq 0$, where $g_{x}(y):=g(x,y)$.

\begin{remark}
\label{RemarkRecoverNoiseless}Of course, if $\omega _{t}=\omega _{0}$ for
all times $t$, then we recover the noiseless case with $f:=f_{\omega _{0}}$, 
$y_{0}=g^{(0)}(x_{0},y_{0})$ and 
\begin{equation}
g_{x_{t-1},...,x_{0}}(y_{0})=g_{x_{t-1},...,x_{1}}(g(x_{0},y_{0}))=g^{(t)}(x_{0},y_{0})
\label{g(t)(x,y) noisy}
\end{equation}
for $t\geq 1$; see equations (\ref{[f,g]n}) and (\ref{g(n)(x,y)}).
\end{remark}

Some basic facts about the noisy driving dynamic $x_{t+1}=f(\omega
_{t},x_{t})$ follow.

\smallskip

\textbf{Fact 1.} Since the parametric sequence $\omega =(\omega _{t})_{t\in 
\mathbb{Z}}$ is a trajectory of an $\Omega $-valued random process modeled
by the shift space $\Sigma =(\Omega ^{\mathbb{Z}},\mathcal{B}^{\times },\mu
,\sigma )$, the noisy orbit%
\begin{eqnarray}
\xi (x,\omega ) &=&(...,x,f_{\omega _{0}}(x),f_{\omega _{1},\omega
_{0}}(x),...,f_{\omega _{t-1},...,\omega _{0}}(x),....)  \label{xi} \\
&=&(...,x_{0},x_{1},x_{2},...,x_{t},...)\in \mathcal{M}_{X}^{\mathbb{Z}} 
\notag
\end{eqnarray}%
is a trajectory of an $\mathcal{M}_{X}$-valued random process. In general,
i.i.d. \negthinspace parametric sequences $\omega $ (commonly used in
applications) do not generate i.i.d. \negthinspace noisy orbits $\xi =\xi
(x,\omega )$. Additive noise $x_{t+1}=f(x_{t})+\omega _{t}$ is a plain
example when the invariant measure of $f:\mathcal{M}_{X}\rightarrow \mathcal{%
M}_{X}$ is not uniform.

\smallskip

\textbf{Fact 2.} Since the $\omega _{t}$'s are the outcomes of a stationary
process, the $x_{t}$'s are also the outcomes of a stationary process.
Indeed, the definition $x_{t+1}=f(\omega _{t},x_{t})$ is time-invariant due
to the stationarity in the generation of the $\omega _{t}$'s.

\smallskip

\textbf{Fact 3.} Under additional assumptions, $\xi (x,\omega
)=(x_{t})_{t\in \mathbb{Z}}$ can match any arbitrary stationary sequence in $%
\mathcal{M}_{X}$ (i.e., a trajectory of a stationary $\mathcal{M}_{X}$%
-valued random process) by fine tuning the sequence $\omega $. For example,
assume the following mild proviso.

\begin{condition}
\label{ConditionO3}$f(\cdot ,x)=f_{x}:\mathcal{M}_{P}\rightarrow \mathcal{M}%
_{X}$ is an embedding for each $x\in \mathcal{M}_{X}$.
\end{condition}

\noindent Under this condition, given $x_{t}$, the relationship between $%
\omega _{t}$ and $x_{t+1}$ is 1-to-1 for each $t$, which implies that the
equation $f_{x_{t}}(\omega _{t}):=f(\omega _{t},x_{t})=x_{t+1}$ can be
solved for $\omega _{t}$ in a unique way. Therefore, the noisy orbit $\xi
(x,\omega )$ can be recursively transformed into any stationary sequence $%
\eta \in \mathcal{M}_{X}^{\mathbb{Z}}$ by choosing $x_{0}=\eta _{0}$ and $%
\omega _{t}$ as the unique solution of $f(\omega _{t},x_{t})=\eta _{t+1}$
for $t=0,1,2,...$ and $t=-1,-2,\ldots $. Henceforth we assume that Condition %
\ref{ConditionO3} is met.

\smallskip

\textbf{Fact 4.} In particular, by Condition \ref{ConditionO3} the
relationship between $\xi =\{x_{t}\}_{t\in \mathbb{Z}}$ and $\{x_{0},\omega
\}$ is 1-to-1, i.e., the function $(x,\omega )\mapsto \xi $ is invertible,
where $\xi _{0}=x$. Therefore, we may indistinctly talk of $x$ and $\omega $%
, or $\xi =(x_{t})_{t\in \mathbb{Z}}$. In practice, one chooses $\omega $ so
that the noisy orbit $\xi $ of $x$ deviates from the noiseless orbit by
small perturbations.

\medskip

By Fact 2, we can view the noisy dynamic (\ref{forcing with dyn noise}) as
stochastic forcing, the compact manifold $\mathcal{M}_{X}$ being the
parameter set and the orbits $\xi =(x_{t})_{t\in \mathbb{Z}}$ of the noisy
driver playing the role of the parameter sequences $\omega =(\omega
_{t})_{t\in \mathbb{Z}}$. This being the case, replace in Theorem \ref%
{ThmStarkOmega} (i) the sequence $\omega \in \Omega ^{\mathbb{Z}}$ with the
noisy orbit $\xi =(x_{t})_{t\in \mathbb{Z}}\in \mathcal{M}_{X}^{\mathbb{Z}}$%
, (ii) the map $f:\Omega \times \mathcal{M}_{X}\rightarrow \mathcal{M}_{X}$
with $g:\mathcal{M}_{X}\times $ $\mathcal{M}_{Y}\rightarrow \mathcal{M}_{Y}$%
, (iii) $f_{\omega _{t}}(x_{t})=f(\omega _{t},x_{t})$ with $%
g_{x_{t}}(y_{t})=g(x_{t},y_{t})$, (iv) $f_{\omega _{t},\omega
_{t-1},..,\omega _{0}}$ with $g_{x_{t},x_{t-1},...,x_{0}}$, and (v) $\varphi
_{X}$ with $\varphi _{Y}$, to derive the following result.

\smallskip

\begin{theorem}
\label{ThmEmbedding(y)noisy}If $\delta \geq 2\dim _{Y}+1$, then there exists
a residual set of $(g,\varphi _{Y})$ such that for any $(g,\varphi _{Y})$ in
this set there is an open dense set of sequences $\xi =(x_{t})_{t\in \mathbb{%
Z}}\in \mathcal{M}_{X}^{\mathbb{Z}}$ such that the map $E_{g,\varphi
_{Y},\xi }:\mathcal{M}_{Y}\rightarrow \mathbb{R}^{\delta }$ defined by 
\begin{equation}
E_{g,\varphi _{Y},\xi }(y)=(\varphi _{Y}(y),\,\varphi
_{Y}(g_{x_{0}}(y)),...,\,\varphi _{Y}(g_{x_{\delta -2},..,x_{0}}(y)))
\label{E(xi,y)}
\end{equation}%
is an embedding.
\end{theorem}

According to equation (\ref{E(xi,y)}), $E_{g,\varphi _{Y},\xi }:\mathcal{M}%
_{Y}\rightarrow \mathbb{R}^{\delta }$ depends actually on the $\delta -1$
parameters $x_{0},x_{1},\ldots ,x_{\delta -2}$. The points $%
(x_{0},...,x_{\delta -2})$ are dense in the finite-dimensional manifold $%
\mathcal{M}_{X}^{\delta -1}$ if and only if the points $x_{k}$ are dense in $%
\mathcal{M}_{X}$ for each $0\leq k\leq \delta -2$. It follows that $%
E_{g,\varphi _{Y},\xi }$ is an embedding for a residual set of $(g,\varphi
_{Y})$ and dense sets of points $\{x_{0},...,x_{\delta -2}\}$ in $\mathcal{M}%
_{X}$.

\begin{remark}
\label{RemarkEmbedding(y)noisy2}One can extend the map $E_{g,\varphi
_{Y},\xi }$ from $\mathcal{M}_{Y}$ to $\mathcal{M}_{X}\times \mathcal{M}_{Y}$
by defining $E_{g,\varphi _{Y},x_{1},\ldots ,x_{\delta -2}}:\{x_{0}\}\times 
\mathcal{M}_{Y}\rightarrow \mathbb{R}^{\delta }$ as 
\begin{equation}
E_{g,\varphi _{Y},x_{1},\ldots ,x_{\delta -2}}(x_{0},y):=E_{g,\varphi
_{Y},\xi }(y).  \label{E_g(x,y)}
\end{equation}%
Yet, $E_{g,\varphi _{Y},x_{1},\ldots ,x_{\delta -2}}$ does not allow to
reconstruct the full state space $\mathcal{M}_{X}\times \mathcal{M}_{Y}$
because, according to Theorem \ref{ThmEmbedding(y)noisy}, in general this
map is an embedding only for a dense set of points $(x_{0},y)\in \mathcal{M}%
_{X}\times \mathcal{M}_{Y}$. Nevertheless, this result can be useful in
applications where one can assume $x_{0}$ to be fixed and generic, like in
time series analysis.
\end{remark}


\section{The noisy scenario}

\label{sec5}

In this section we discuss some changes and limitations introduced by noise
in the conventional framework of Section \ref{sec2}. Since the driver
dynamic now explicitly depends on time through the noise, so do the main
concepts like state reconstruction, cross map and synchronization map.

\subsection{State reconstruction}

Let $\omega \in \Omega ^{\mathbb{Z}}$ be a parametric sequence and suppose $%
d\geq 2\dim _{X}+1$. Then, according to Theorem \ref{ThmStarkOmega}, the map 
$E_{f,\varphi _{X},\omega }:\mathcal{M}_{X}\rightarrow \mathbb{R}^{d}$
defined in equation (\ref{F omega}) is generically an embedding. Similarly
to the noiseless case, define the manifolds 
\begin{equation}
\mathcal{N}_{X,\omega }=E_{f,\varphi _{X},\omega }(\mathcal{M}_{X})\subset 
\mathbb{R}^{d}  \label{N_X,omega}
\end{equation}%
(each one diffeomorphic to $\mathcal{M}_{X}$) and the noisy\ time delay
vectors 
\begin{eqnarray}
\mathbf{x}_{t}&=&E_{f,\varphi _{X},\sigma ^{t}(\omega )}(x_{t}) \label{x_t noisy} \\
&=&(\varphi_{X}(x_{t}),\,\varphi _{X}(x_{t+1}),...,\,\varphi _{X}(x_{t+d-1}))\in 
\mathcal{N}_{X,\sigma ^{t}(\omega )}  \nonumber
\end{eqnarray}
with $x_{t+k}=f_{\omega _{t+k-1},...,\omega _{t}}(x_{t})$ for $k\geq 1$.
Then, the driver dynamics $x_{t+1}=f_{\omega _{t}}(x_{t})$ translate to 
\begin{equation}
\mathbf{x}_{t+1}=F_{\sigma ^{t}(\omega )}(\mathbf{x}_{t})  \label{F(x_t)}
\end{equation}%
in the reconstructed state spaces, where the map $F_{\omega }:\mathcal{N}%
_{X,\omega }\rightarrow \mathcal{N}_{X,\sigma (\omega )}$ defined as%
\begin{equation}
F_{\omega }=E_{f,\varphi _{X},\sigma (\omega )}\circ f_{\omega _{0}}\circ
E_{f,\varphi _{X},\omega }^{-1}  \label{F_omega}
\end{equation}%
is a diffeomorphism, provided that $E_{f,\varphi _{X},\omega }$ and $%
E_{f,\varphi _{X},\sigma (\omega )}$ are embeddings. At variance with the
noiseless case, the reconstructed dynamic $\mathbf{x}_{t}\mapsto \mathbf{x}%
_{t+1}$ hops from a diffeomorphic copy $\mathcal{N}_{X,\sigma ^{t}(\omega )}$
of $\mathcal{M}_{X}$ to another diffeomorphic copy $\mathcal{N}_{X,\sigma
^{t+1}(\omega )}$.

Likewise, let $\xi =\xi (x,\omega )\in \mathcal{M}_{X}^{\mathbb{Z}}$ be a
noisy orbit of $x=x_{0}$ (equation (\ref{xi})) and suppose $\delta \geq
2\dim _{Y}+1$. Then, according to Theorem \ref{ThmEmbedding(y)noisy}, the
map $E_{g,\varphi _{Y},\xi }:\mathcal{M}_{Y}\rightarrow \mathbb{R}^{\delta }$
defined in equation (\ref{E(xi,y)}) is generically an embedding. Define the
manifolds 
\begin{equation}
\mathcal{N}_{Y,\xi }=E_{g,\varphi _{Y},\xi }(\mathcal{M}_{Y})\subset \mathbb{%
R}^{\delta }  \label{N_Y,omega}
\end{equation}%
(each one diffeomorphic to $\mathcal{M}_{Y}$) and the noisy\ time delay
vectors 
\begin{eqnarray}
\mathbf{y}_{t}&=&E_{g,\varphi _{Y},\sigma ^{t}(\xi )}(y_{t}) \label{y_t noisy} \\
&=&(\varphi_{Y}(y_{t}),\,\varphi _{Y}(y_{t+1}),...,\,\varphi _{Y}(y_{t+\delta -1}))\in 
\mathcal{N}_{Y,\sigma ^{t}(\xi )},  \nonumber
\end{eqnarray}%
with $y_{t+1}=g(x_{t},y_{t})=:g_{x_{t}}(y_{t})$ and 
\begin{equation}
y_{t+k}=g(x_{t+k-1},g_{x_{t+k-2},...,x_{t+1},x_{t}}(y_{t}))=:g_{x_{t+k-1},...,x_{t+1},x_{t}}(y_{t})
\label{y_(t+k)=g(k)}
\end{equation}%
for $k\geq 2$. Then, similarly to (\ref{F_omega}), the map $G_{\xi }:%
\mathcal{N}_{Y,\xi }\rightarrow \mathcal{N}_{Y,\sigma (\xi )}$ defined as%
\begin{equation}
G_{\xi }=E_{g,\varphi _{Y},\sigma (\xi )}\circ g_{\xi _{0}}\circ
E_{g,\varphi _{Y},\xi }^{-1}  \label{G_xi}
\end{equation}%
is a diffeomorphism, provided that $E_{g,\varphi _{y},\xi }$ and $%
E_{g,\varphi _{y},\sigma (\xi )}$ are embeddings, and it holds%
\begin{equation}
\mathbf{y}_{t+1}=G_{\sigma ^{t}(\xi )}(\mathbf{y}_{t}).  \label{G(y_t)}
\end{equation}%
Again, the range of $G_{\sigma ^{t}(\xi )}$ depends on $t$ through $\sigma
^{t}(\xi )$, but all of them are diffeomorphic copies of $\mathcal{M}_{Y}.$


\subsection{Cross map}

\label{sec5.1}

The definition of the cross map of the systems $X\rightsquigarrow Y$ in (\ref%
{Phi}) hinges on the reconstruction of both the driver state space $\mathcal{%
M}_{X}$ and the full state space $\mathcal{M}_{X}\times \mathcal{M}_{Y}$.
However, according to Remark \ref{RemarkEmbedding(y)noisy2}, the latter
reconstruction is generally only possible in the noisy scenario for a dense
set of driver states.

This being the case, we are going to define the cross map $\mathbf{x}%
_{t}=\Phi _{\sigma ^{t}(\omega )}(\mathbf{y}_{t})$ for two time series $(%
\mathbf{x}_{t})_{t\geq 0}$ and $(\mathbf{y}_{t})_{t\geq 0}$ of time-delay
vectors obtained from a noisy orbit $\xi =\xi (x_{0},\omega )$ of the driver 
$X$ (equation (\ref{xi})) and the corresponding response from the system $Y$%
, respectively. This limited approach suffices for the needs of time series
analysis, where the focus in practice is on (finite segments of) single
orbits rather than on manifolds, and points and parameters can be considered
generic. For ease of notation, we will write 
\begin{equation}
E_{g,\varphi _{Y},\sigma ^{t}(\omega )}(x_{t},y_{t})=E_{g,\varphi
_{Y},x_{t+1},\ldots ,x_{t+\delta -2}}(x_{t},y_{t})=\mathbf{y}_{t},
\label{y_t noisy 2}
\end{equation}%
see equation (\ref{E_g(x,y)}), since the relation $\sigma ^{t}(\xi
)\leftrightarrow (x_{t},\sigma ^{t}(\omega ))$ is 1-to-1 by Condition \ref%
{ConditionO3} in Section \ref{sec4}.

To define the cross map in the presence of dynamical noise, $\mathbf{x}%
_{t}=\Phi _{\sigma ^{t}(\omega )}(\mathbf{y}_{t})$, we mimic the definition
of the cross map in equation (\ref{chain}) in the form%
\begin{equation}
\begin{array}{rcccl}
\mathcal{M}_{X}\times \mathcal{M}_{Y}\ni & (x_{t},y_{t}) & \overset{\Pi _{X}}%
{\longrightarrow } & x_{t} & \in \mathcal{M}_{X} \\ 
E_{g,\varphi _{Y},\sigma ^{t}(\omega )}^{-1} & \uparrow &  & \downarrow & 
E_{f,\varphi _{X},\sigma ^{t}(\omega )} \\ 
\mathcal{N}_{Y,\sigma ^{t}(\xi )}\ni & \mathbf{y}_{t} &  & \mathbf{x}_{t} & 
\in \mathcal{N}_{X,\sigma ^{t}(\omega )}%
\end{array}
\label{noisy chain}
\end{equation}%
under the assumption that $E_{g,\varphi _{Y},\sigma ^{t}(\omega
)}(x_{t},y_{t})$ is an embedding for the considered states $x_{t}\in 
\mathcal{M}_{X}$. Hence, 
\begin{equation}
\mathbf{x}_{t}=\Phi _{\sigma ^{t}(\omega )}(\mathbf{y}_{t}):=(E_{f,\varphi
_{X},\sigma ^{t}(\omega )}\circ \Pi _{X}\circ E_{g,\varphi _{Y},\sigma
^{t}(\omega )}^{-1})(\mathbf{y}_{t}).  \label{noisy cross map}
\end{equation}

Let us check that $\Phi _{\sigma ^{t}(\omega )}(\mathbf{y}_{t})$ becomes $%
\Phi (\mathbf{y}_{t})$, equation (\ref{Phi}), when the noise is switched off
in equation (\ref{noisy cross map}), i.e., when $\omega =\bar{\omega}$ with $%
\bar{\omega}_{t}=\omega _{0}$ for all $t\in \mathbb{Z}$. We suppose that the
maps $E_{f,\varphi _{X},\sigma ^{t}(\omega )}$ and $E_{g,\varphi _{Y},\sigma
^{t}(\omega )}$ are embeddings for $\omega =\bar{\omega}$.

In that case, $E_{f,\varphi _{X},\sigma ^{t}(\bar{\omega})}(\mathbf{x}%
_{t})=E_{f,\varphi _{X}}(\mathbf{x}_{t})$ with $f:=f_{\omega _{0}}$; see
equations (\ref{x_t noisy}) and (\ref{F}). Similarly, by equations (\ref%
{g(t)(x,y) noisy}) and (\ref{G}) with $D=\delta -1$, and setting $\bar{\xi}%
=(x_{0},\bar{\omega})=(f^{t}(x_{0}))_{t\in \mathbb{Z}}$, 
\begin{eqnarray}
&&E_{g,\varphi _{Y},\sigma ^{t}(\bar{\omega})}(x_{t},y_{t}) \label{G noiseless} \\
&=&E_{g,\varphi_{Y},\sigma ^{t}(\bar{\xi})}(y_{t})  \notag \\
&=&(\varphi _{Y}(y_{t}),\,\varphi _{Y}(g_{x_{t}}(y_{t})),...,\,\varphi
_{Y}(g_{x_{t+\delta -2},..,x_{t}}(y_{t})))  \notag \\
&=&(\varphi _{Y}(g^{(0)}(x_{t},y_{t})),\,\varphi
_{Y}(g^{(1)}(x_{t},y_{t})),...,\,\varphi _{Y}(g^{(\delta -1)}(x_{t},y_{t})))
\notag \\
&=&E_{f,g,\varphi _{Y}}(x_{t},y_{t}).  \notag
\end{eqnarray}%
Comparison with equation (\ref{Phi}) shows that $\Phi _{\sigma ^{t}(\bar{%
\omega})}(\mathbf{y}_{t})=\Phi (\mathbf{y}_{t})$, as it should.


\subsection{Synchronization map}

\label{sec5.2}

Generalized synchronization (\ref{sync map}) can be extended from the
noiseless dynamic $[f,g]$ to the noisy dynamic $[f_{\omega },g]$, $\omega
\in \Omega ^{\mathbb{Z}}$, where the dynamic changes at every time step, by
requiring 
\begin{equation}
y_{t}=h_{\sigma ^{t}(\omega )}(x_{t}).  \label{y=h_om(x)}
\end{equation}

\begin{definition}
We say that the responder $Y$ is synchronized to a driver $X$ perturbed by
the noise $\omega \in \Omega ^{\mathbb{Z}}$, if there is a sequence of
continuous maps $h_{\sigma ^{t}(\omega )}:\mathcal{M}_{X}\rightarrow 
\mathcal{M}_{Y}$ such that equation (\ref{y=h_om(x)}) holds for all $t\in 
\mathbb{Z}$.
\end{definition}

More generally, the synchronization map of order $k\geq 0$, $h^{(k)}=h\circ
f^{k}$ (equation (\ref{sync map B})), generalizes to $h_{\sigma ^{t}(\omega
)}^{(0)}:=h_{\sigma ^{t}(\omega )}$ and 
\begin{equation}
y_{t}=h_{\sigma ^{t}(\omega )}^{(k)}(x_{t-k}):=h_{\sigma ^{t}(\omega )}\circ
f_{\omega _{t-1},...,\omega _{t-k}}(x_{t-k})  \label{noisy sync map order k}
\end{equation}%
for $k\geq 1$ in the noisy case, while the synchronization map of period $%
K\geq 1$ (\ref{multiv h}) generalizes to%
\begin{equation}
y_{t}=\frac{1}{K}\sum_{k=0}^{K-1}h_{\sigma ^{t}(\omega
)}^{(k)}(x_{t-k})=:h_{K,\sigma ^{t}(\omega )}(x_{t},...,x_{t-K+1}).
\label{noisy multiv map K}
\end{equation}

To define the syncronization map in the reconstructed spaces, we replace $%
\mathbf{y}_{t}=\Phi ^{-1}(\mathbf{x}_{t})$ with $\mathbf{y}_{t}=H_{\sigma
^{t}(\omega )}(\mathbf{x}_{t})$ in the \textquotedblleft
noisy\textquotedblright\ version of diagram (\ref{diagram}):%
\begin{equation}
\begin{array}{rcccl}
\mathcal{M}_{X}\ni & x_{t} & \overset{h_{\sigma ^{t}(\omega )}}{%
\longrightarrow } & y_{t} & \in \mathcal{M}_{Y} \\ 
E_{f,\varphi _{X},\sigma ^{t}(\omega )}^{-1} & \uparrow &  & \uparrow & 
E_{g,\varphi _{Y},\sigma ^{t}(\xi )}^{-1} \\ 
\mathcal{N}_{X,\sigma ^{t}(\omega )}\ni & \mathbf{x}_{t} & \overset{%
H_{\sigma ^{t}(\omega )}}{\longrightarrow } & \mathbf{y}_{t} & \in \mathcal{N%
}_{Y,\sigma ^{t}(\xi )}%
\end{array}
\label{diagram noisy}
\end{equation}%
Here we used equations (\ref{x_t noisy}) and (\ref{y_t noisy}), and assume
that the maps $E_{f,\varphi _{X},\sigma ^{t}(\omega )}$ and $E_{g,\varphi
_{Y},\sigma ^{t}(\xi )}$ are embeddings. Then it follows from (\ref{diagram
noisy}) 
\begin{eqnarray}
\mathbf{y}_{t}&=&(E_{g,\varphi _{Y},\sigma ^{t}(\xi )}\circ h_{\sigma
^{t}(\omega )}\circ E_{f,\varphi _{X},\sigma ^{t}(\omega )}^{-1})(\mathbf{x}%
_{t}) \nonumber \\
&=:&H_{\sigma ^{t}(\omega )}(\mathbf{x}_{t}).  \label{H_omega(x,t)}
\end{eqnarray}

Furthermore, by equation (\ref{F(x_t)}),%
\begin{eqnarray}
\mathbf{x}_{t} &=&F_{\sigma ^{t-1}(\omega )}(\mathbf{x}_{t-1})=...=(F_{%
\sigma ^{t-1}(\omega )}\circ ...\circ F_{\sigma ^{t-k}(\omega )})(\mathbf{x}%
_{t-k})  \notag \\
&=&F_{\sigma ^{t-1}(\omega )}^{(k)}(\mathbf{x}_{t-k}),  \label{x=F(k)(x)}
\end{eqnarray}%
so that%
\begin{eqnarray}
\mathbf{y}_{t}&=&\frac{1}{K}\sum\limits_{k=0}^{K-1}\left( H_{\sigma
^{t}(\omega )}\circ F_{\sigma ^{t-1}(\omega )}^{(k)}\right) (\mathbf{x}%
_{t-k}) \nonumber \\
&=:&H_{K,\sigma ^{t}(\omega )}(\mathbf{x}_{t},\ldots ,\mathbf{x}%
_{t-K+1})  \label{mult syn map noisy}
\end{eqnarray}%
generalizes the reconstructed synchronization map of period $K$, equation (%
\ref{multi h reconst 2}), to the noisy case.

To check that $\mathbf{y}_{t}=H_{\sigma ^{t}(\omega )}(\mathbf{x}_{t})$
becomes $\mathbf{y}_{t}=\Phi ^{-1}(\mathbf{x}_{t})$ when the noise is
switched off in equation (\ref{H_omega(x,t)}), replace 
\begin{equation}
E_{g,\varphi _{Y},\sigma ^{t}(\xi )}^{-1}:\mathbf{y}_{t}\longrightarrow y_{t}
\label{Eq  aux}
\end{equation}%
on the right column of diagram (\ref{diagram noisy}) with%
\begin{equation}
\Pi _{Y}\circ E_{g,\varphi _{Y},\sigma ^{t}(\omega )}^{-1}:\mathbf{y}_{t}%
\overset{E_{g,\varphi _{Y},\sigma ^{t}(\omega )}^{-1}}{\longrightarrow }%
(x_{t},y_{t})\overset{\Pi _{Y}}{\longrightarrow }y_{t}  \label{Eq aux 2}
\end{equation}%
so that, according to equation (\ref{G noiseless}), $\Pi _{Y}\circ
E_{g,\varphi _{Y},\sigma ^{t}(\omega )}^{-1}(\mathbf{y}_{t})$ becomes $\Pi
_{Y}\circ E_{f,g,\varphi _{Y}}^{-1}(\mathbf{y}_{t})$ in the noiseless case $%
\omega =\bar{\omega}$, i.e. $\bar{\omega}_{t}=\omega _{0}$ for all $t\in 
\mathbb{Z}$. Finally, set $h(x_{t})=h_{\sigma ^{t}(\bar{\omega})}$ to
convert diagram (\ref{diagram noisy}) to diagram (\ref{diagram}), thus
identifying $H_{\sigma ^{t}(\bar{\omega})}(\mathbf{x}_{t})$ with $\Phi ^{-1}(%
\mathbf{x}_{t})$, as it should be.

The numerical simulations of Section \ref{sec6} show that synchronization is
robust against dynamical noise for strong enough couplings and, hence, can
occur in the presence of dynamical noise. On the other hand, if
synchronization occurs in the presence of dynamical noise but disappears
when noise is switched off, then one speaks of noise-induced synchronization 
\cite{Moskalenko2011}.

Finally, we can generalize the concepts of asymptotic synchronization and
stability of the responder in the presence of noise in the driver as
follows. We say that $Y$ is aymptotically synchronized to the noisy driver $%
X $ if the definition of synchronization, equation (\ref{H_omega(x,t)}),
holds only asymptotically, i.e., 
\begin{equation}
\lim_{t\rightarrow \infty }\left\Vert \mathbf{y}_{t}-H_{\sigma ^{t}(\omega
)}(\mathbf{x}_{t})\right\Vert =0,  \label{noisy asynp sync}
\end{equation}%
where $\left\Vert \cdot \right\Vert $ is a norm in $\mathbb{R}^{\dim _{Y}}$.
It follows then that $Y$ is \textit{asymptotically stable}, i.e., the orbits
of $Y$ converge to $H_{\sigma ^{t}(\omega )}(\mathbf{x}_{t})$ regardless of
their initial conditions. Asymptotic stability can easily be checked in
practice. As in the noisless case, it is a handy method to rule out
synchronization.


\section{Numerical simulations}

\label{sec6}

Unlike identical synchronization, which can be easily visualized,
generalized synchronization is more difficult to detect. As mentioned in the
Introduction, there exists an extensive literature on methods to detect
functional dependency (and generalized synchronization for that matter)
between two time series. The functional dependency targeted in this section
is the synchronization map of a certain period $K>1$ given in equations (\ref%
{multi h reconst 2}) and (\ref{mult syn map noisy}) for noiseless and noisy
drivers, respectively. For this reason, we use recurrent neural networks of
the type \textit{long short-term memory} (LSTM), which excel at predicting
data from time series and are robust to noise. In fact, the LSTM nets
outperformed the perceptrons ($K=1$) in the numerical simulations below, so
we will only report the results obtained with the former. As a benchmark we
use nearest-neighbor cross prediction (Section \ref{sec2.2}) because it is
based on the continuity of the cross map (and its inverse in case of
synchronization). In addition, nearest-neighbor cross prediction is robust
against noise, particularly if the neighborhoods are well populated.

\subsection{Models}

\label{sec6.1}

For the numerical simulations we chose two unidirectionally coupled H\'{e}%
non maps with several structural parameters and varying coupling strength.
This testbed, first proposed by Schiff et al. \cite{Schiff1996} and studied with the
normalized mutual error, has been revisited several times in the literature,
e.g., in Quian Quiroga et al. \cite{QuianQuiroga2000}, where the authors use the conditional
Lyapunov exponent and the so-called nonlinear interdependencies \cite%
{Arnhold1999}.

Thus, the equations of the driver $X$, with states $x=(x^{(1)},x^{(2)})$ in
a trapping region of the attractor, are%
\begin{equation}
\left\{ 
\begin{array}{l}
x_{t+1}^{(1)}=1.4-(x_{t}^{(1)}{})^{2}+(b_{1}+\omega _{t})x_{t}^{(2)} \\ 
x_{t+1}^{(2)}=x_{t}^{(1)}%
\end{array}%
\right.  \label{HenonMapX}
\end{equation}%
where $b_{1}$ is a constant and $\omega _{t}$ are i.i.d. \negthinspace
random numbers in the interval $[-A,A]$, the noiseless scenario
corresponding to $A=0$. The observation function is $\varphi
_{X}(x_{t})=x_{t}^{(1)}$, i.e., the projection on the first component.

The equations of the responder $Y$, with states $y=(y^{(1)},y^{(2)})$, are 
\begin{equation}
\left\{ 
\begin{array}{l}
y_{t+1}^{(1)}=1.4-[Cx_{t}^{(1)}y_{t}^{(1)}+(1-C)(y_{t}^{(1)}{})^{2}]+b_{2}y_{t}^{(2)}
\\ 
y_{t+1}^{(2)}=y_{t}^{(1)}%
\end{array}%
\right.  \label{HenonMap2}
\end{equation}%
where $b_{2}$ is a constant and $C$ is the \textit{coupling strength}. For $%
C=0$, systems $X$ and $Y$ are uncoupled. The observation function is again
the projection on the first component, $\varphi _{Y}(y_{t})=y_{t}^{(1)}$.

The parameter settings are as follows.

\begin{itemize}
\item The settings for the constants $b_{1}$ and $b_{2}$ are the same as 
in Schiff et al. \cite{Schiff1996} and 
Quian Quiroga et al. \cite{QuianQuiroga2000}. So, we first set $b_{1}=b_{2}=0.3$, the
standard values of the H\'{e}non map, to study the coupling of identical
systems (\textit{Model H\'{e}non 0.3-0.3}), which allows identical
synchronization (i.e., $y_{t}=x_{t}$) for $C=1$. Then, to study the coupling
of non-identical systems, we set $b_{1}=0.3$, $b_{2}=0.1$ (\textit{Model H%
\'{e}non 0.3-0.1}) and $b_{1}=0.1$, $b_{2}=0.3$ (\textit{Model H\'{e}non
0.1-0.3}).

\item For the previous choices of $b_{1}$ and $b_{2}$ we found that the
driver orbits can diverge for noise amplitudes\textbf{\ }$A>0.013$\textbf{, }%
so we restrict them to the interval $0\leq A\leq 0.013$\textbf{. }The
amplitudes used in the figures below are $A=0$ (noiseless driver), $0.005$
and $0.013$.

\item The range of the \textit{coupling strength} $C$ is $0\leq C\leq 1.2$;\
the increment of $C$ in the figures below is $\Delta C=0.05$.

\item For each case described above (identical/non-identical systems,
noiseless/noisy driver), one series $(x_{t})_{0\leq t\leq T-1}$ and one
series $(y_{t})_{0\leq t\leq T-1}$ were generated with seeds $x_{0}=(0,0.9)$
and $y_{0}=(0.75,0)$, and length $T=10^{5}$ (after discarding the first $%
1000 $ points). Since we are only interested in synchronization, one series
per case suffices because of asymptotic stability (Section \ref{sec2.3}).

\item The embedding dimension in the noiseless and noisy scenarios is $d=5$,
i.e., $\mathbf{x}_{t}=(x_{t}^{(1)},...,x_{t+4}^{(1)})$ and $\mathbf{y}%
_{t}=(y_{t}^{(1)},...,y_{t+4}^{(1)})$, $0\leq t\leq T-5$. A posteriori
justification for this choice are the excellent results obtained in the
benchmark below.
\end{itemize}

The methods to test for synchronization in the \textit{noiseless case} ($A=0$%
) and \textit{noisy cases} ($A=0.005$, $0.013$) are the following.

\begin{description}
\item[Method 1] Our first method unveils synchronization by detecting
functional dependencies, namely, the existence of the synchronization map of
period $K=10$ for time delay vectors, i.e., 
\begin{equation}
\mathbf{y}_{t}=H_{10}(\mathbf{x}_{t},\mathbf{x}_{t-1},...,\mathbf{x}_{t-9})
\label{multi h tilde K=10}
\end{equation}%
(see equation (\ref{multi h reconst 2})). To do this, we used a 3-layer
neural network to predict $\mathbf{y}_{t}$ based on $\mathbf{x}_{t},...,%
\mathbf{x}_{t-9}$. Specifically: (i) the input layer consisted of an LSTM
net with $5$ units, hidden states of dimension $10$ (corresponding to the
inputs $\mathbf{x}_{t},...,\mathbf{x}_{t-9}$) and the activation function $%
\mathrm{ReLU}(x)=\max \{0,x\}$; (ii) the intermediate layer had $25$ neurons
and the activation function $\mathrm{Sigmoid}(x)=1/(1+e^{-x})$; and (iii)
the output layer had $5$ neurons. Hence, the output layer returns $5$
states, corresponding to the $5$ components of $\mathbf{\hat{y}}_{t}$, the
prediction of $\mathbf{y}_{t}$. The network was trained with an $80\%$ of
the data (the first $80,000$ time-delay vectors) and stochastic gradient
descend, while the remaining $20\%$ of the data was used for testing. The
accuracy of the predictions $\mathbf{\hat{y}}_{t}$ output by the neural
network based on the \textit{testing} data $\mathbf{x}_{t}$, ..., $\mathbf{x}%
_{t-9}$ (i.e., for each $t=80,000$, ..., $99,990$) was measured by their
mean squared error (MSE), mean absolute error (MAE) and mean absolute scaled
error (MASE). The matching of these three metrics in both the training and
testing phases discards overfitting. Furthermore, since predictions based on
data patterns are robust against low levels of noise, we expect this method
to work well in both the noiseless and noisy cases.

\item[Method 2] As a benchmark we used nearest-neighbor cross prediction
which, in the noiseless case,\ estimates $\mathbf{x}_{t}$ based on the
continuity of the cross map $\mathbf{x}_{t}=\Phi (\mathbf{y}_{t})$ and
corresponding nearest neighbors \cite{LeVanQuyen1999}. Following the
convergent cross mapping (CCM) method, we measured the accuracy of those
estimations by $r(\mathbf{x},\mathbf{\hat{x}})$, the Pearson correlation
coefficient of the estimates $\mathbf{\hat{x}}_{t}$ obtained with the $d+1=6$
nearest neighbors of $\mathbf{y}_{t}$. Since $T=10^{5}$, the time series are
sufficiently long to obtain good estimates (actually only the first $10,000$
points were used), so $r(\mathbf{x},\mathbf{\hat{x}})\simeq 1$. On the
contrary, if $r(\mathbf{y},\mathbf{\hat{y}})$ is the Pearson correlation
coefficient of the estimates $\mathbf{\hat{y}}_{t}$ obtained via the $d+1=6$
nearest-neighbors of $\mathbf{x}_{t}$, then we expect $r(\mathbf{y},\mathbf{%
\hat{y}})\simeq 0$, unless $Y$ synchronizes with $X$, in which case $\mathbf{%
y}=\Phi ^{-1}(\mathbf{x})$ and $r(\mathbf{y},\mathbf{\hat{y}})\simeq 1$
(this time due to the continuity of $\Phi ^{-1}$). We conclude that, if 
\begin{equation}
\Delta r=r(\mathbf{x},\mathbf{\hat{x}})-r(\mathbf{y},\mathbf{\hat{y}})
\label{Delta rho}
\end{equation}%
and $X\rightsquigarrow Y$, then

\begin{description}
\item[(i)] $0\leq $ $\Delta r\leq 1$, and

\item[(ii)] $\Delta r\simeq 0$ signalizes synchronization, except when $r(%
\mathbf{x},\mathbf{\hat{x}})=0=r(\mathbf{y},\mathbf{\hat{y}})$, i.e., when $%
X $ and $Y$ are uncoupled.
\end{description}

In the noisy cases, the situation is qualitatively the same thanks to the
robustness of nearest-neighbor cross prediction against low levels of noise.
See, e.g., Sugihara et al. \cite{Sugihara2012},  M\o nster et al. \cite{Monster2017} 
and the book by Datseris and Parlitz \cite{Datseris2022} for CCM algorithms
to compute (\ref{Delta rho}).
\end{description}


\subsection{Results}

\label{sec6.2}

Out of the accuracy results obtained with the LSTM network and testing data,
we are going to discuss only the MSE vs $C$ curves since the other two
curves, MAE and MASE vs $C$, are similar for all models.

\begin{figure*}[tbh]
\begin{center}
\includegraphics[width=120mm]{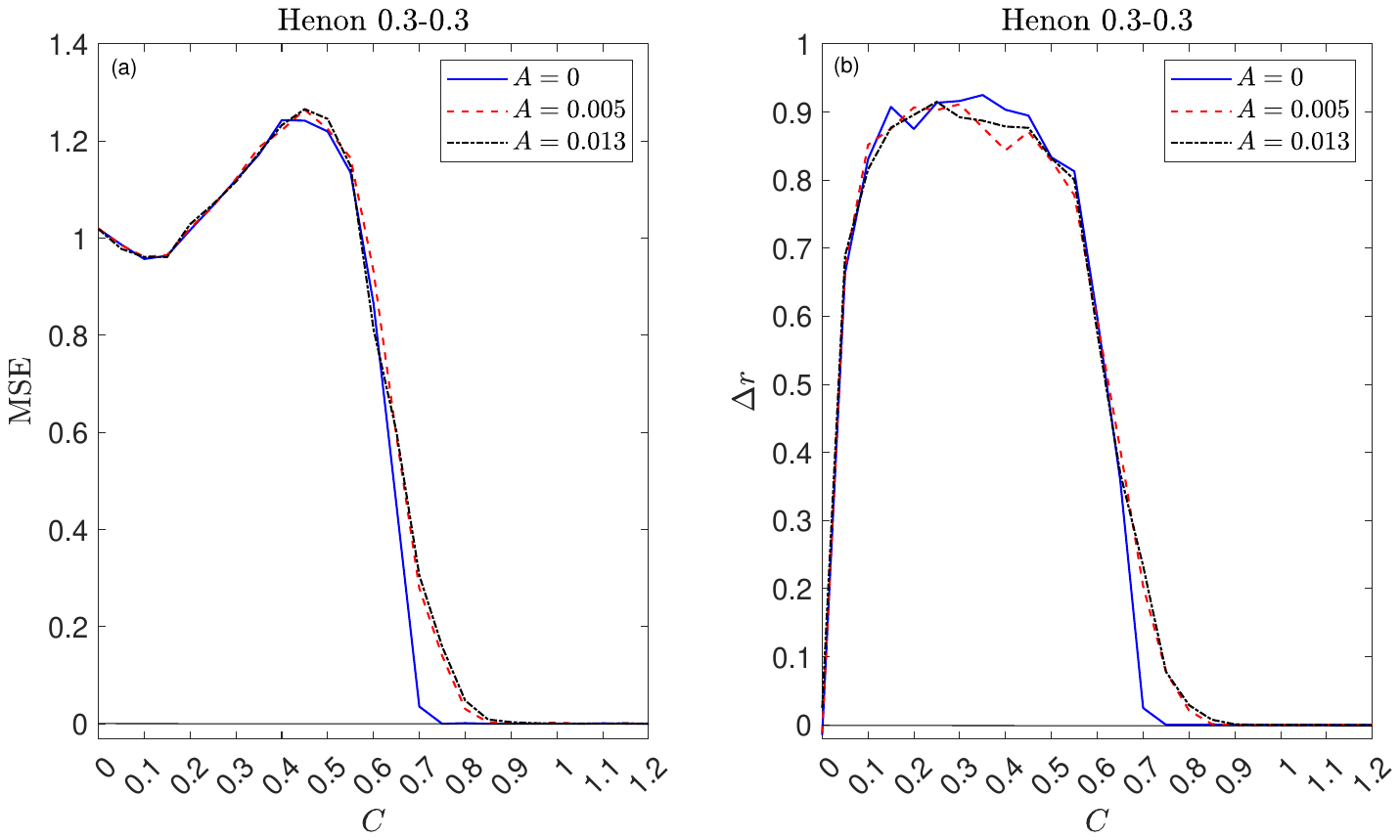}
\caption{ Numerical results for the model H\'{e}non 0.3-0.3, i.e., $%
b_{1}=0.3 $ in (\protect\ref{HenonMapX}) and $b_{2}=0.3$ in (\protect\ref%
{HenonMap2}). (a) \textrm{MSE} vs the coupling strength $C$ for a noiseless
driver (noise amplitude $A=0$) and a noisy driver ($A=0.005$, $0.013$)
obtained with an LSTM net. (b) $\Delta r$ vs $C$ for a noiseless driver ($%
A=0 $) and a noisy driver ($A=0.005$, $0.013$) obtained via $6$%
-nearest-neighbor cross prediction. See text for more detail.}
\label{figure1}
\end{center}
\end{figure*}

\begin{figure*}[th]
\begin{center}
\includegraphics[width=150mm]{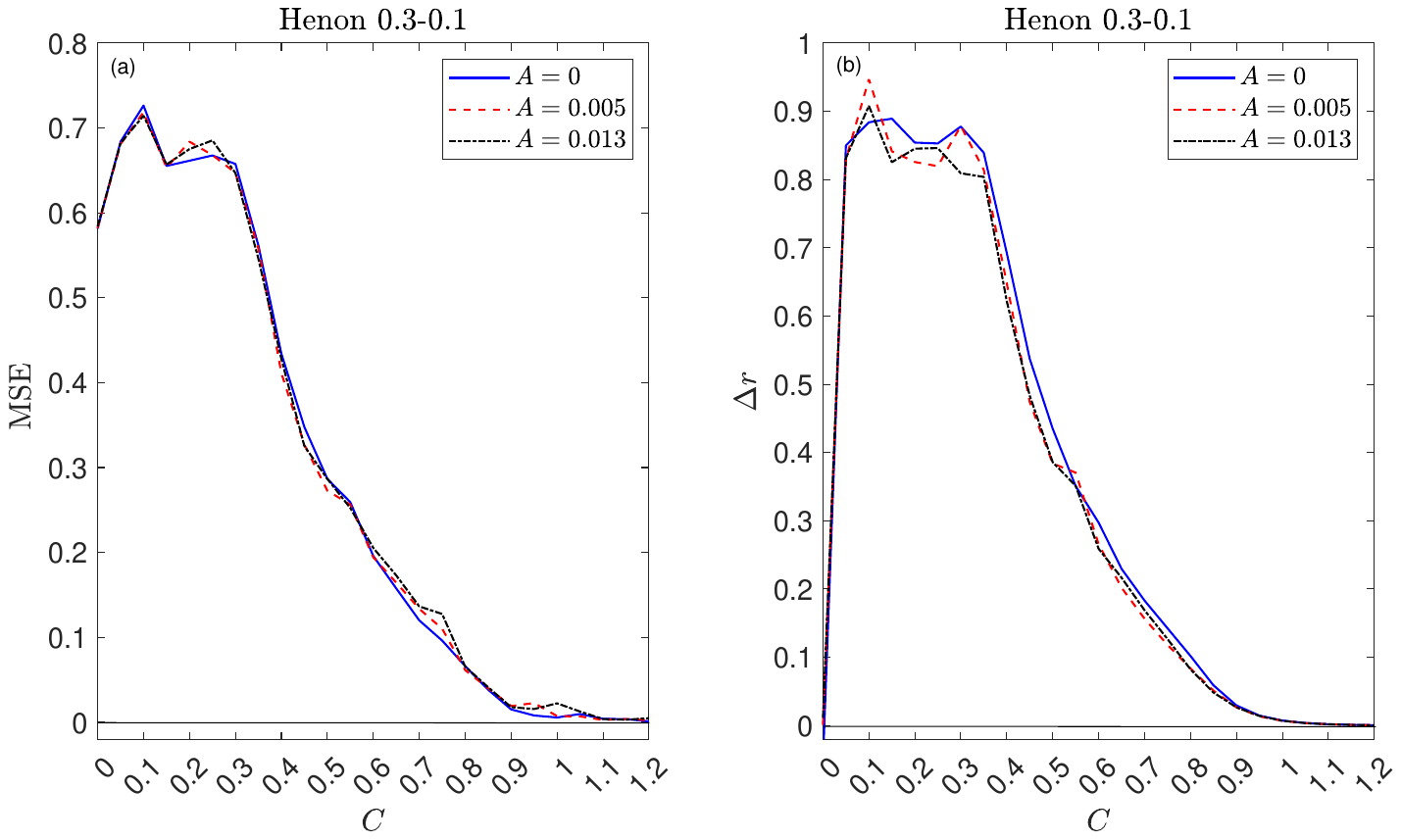}
\caption{ Numerical results for the model H\'{e}non 0.3-0.1, i.e., $%
b_{1}=0.3 $ in (\protect\ref{HenonMapX}) and $b_{2}=0.1$ in (\protect\ref%
{HenonMap2}). The information displayed in the panels (a) and (b) is the
same as in Figure \protect\ref{figure1}.}
\label{figure2}
\end{center}
\end{figure*}

\begin{figure*}[th]
\begin{center}
\includegraphics[width=150mm]{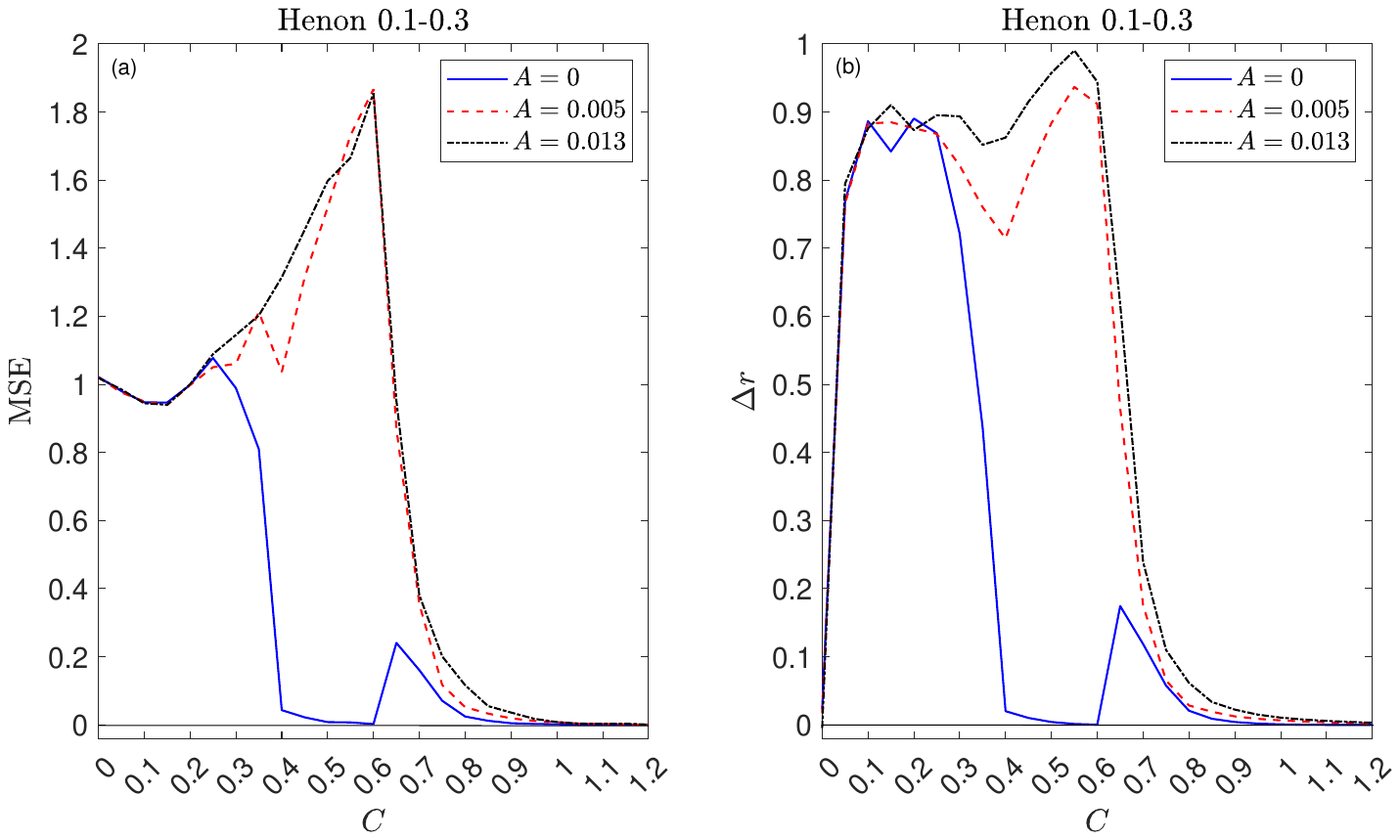}
\caption{ Numerical results for the model H\'{e}non 0.1-0.3, i.e., $%
b_{1}=0.1 $ in (\protect\ref{HenonMapX}) and $b_{2}=0.3$ in (\protect\ref%
{HenonMap2}). The information displayed in the panels (a) and (b) is the
same as in Figure \protect\ref{figure1}.}
\label{figure3}
\end{center}
\end{figure*}

The results of the numerical simulations are depicted in Figures \ref%
{figure1}-\ref{figure3} for Method 1 (panels (a)) and Method 2 (panels (b))
and the three models H\'{e}non 0.3-0.3 (Figure \ref{figure1}), 0.3-0.1
(Figure \ref{figure2}), and 0.1-0.3 (Figure \ref{figure3}). Comparison of
both panels for each case and $C>0$ shows an excellent agreement of both
methods on the synchronization states, i.e., \textrm{MSE}$(C)=0$ in panels
(a) and $\Delta r(C)=0$ in panels (b). As noted above, $\Delta r(0)=0$ in
all cases owing to the fact that $X$ and $Y$ are uncoupled for $C=0$; such
numerical artefacts can be easily filtered out by checking whether $r(%
\mathbf{x},\mathbf{\hat{x}})\simeq 0$ and $r(\mathbf{y},\mathbf{\hat{y}}%
)\simeq 0.$

The main conclusions from the numerical results can be summarized as follows.

\begin{itemize}
\item Small-amplitude noise does not destroy all the states of
\textquotedblleft strong\textquotedblright\ synchronization (i.e., due to
strong enough couplings) but only shifts the synchronization threshold to
higher values. So, synchronization can also occur in the presence of
dynamical noise.

\item Synchronization due to strong enough couplings is robust against
small-amplitude dynamical noise, while synchronization states with a weak
coupling strength can be unstable whatever the amplitude of the noise. This
fact is illustrated in the Model H\'{e}non 0.1-0.3 (Figure \ref{figure3}),
where synchronization is detected in the interval $0.5\lesssim C\lesssim 0.6$
for $A=0$.

\item Weakly coupled systems can be asymptotically synchronized, which can
be detected via the auxiliary systems method both in the noiseless and noisy
cases. Indeed, Table \ref{Table asymp stability} shows the intervals of
coupling strengths for which the responder is asymptotically stable,
obtained with the auxiliary system method. Therefore, synchronization can
occur only for couplings in the corresponding interval (as it does). Note
that Table \ref{Table asymp stability} excludes the spurious synchronization 
$\Delta r(0)=0$. 
\begin{table}[tbh]
\par
\begin{center}
\begin{tabular}{|l|c|c|c|}
\hline
& \textbf{H\'{e}non 0.3-0.3} & \textbf{H\'{e}non 0.3-0.1} & \textbf{H\'{e}%
non 0.1-0.3} \\ \hline
$A=0$ & $0.40\leq C\leq 1.20$ & $0.15\leq C\leq 1.20$ & $0.40\leq C\leq 1.20$
\\ 
$A=0.005$ & $0.40\leq C\leq 1.20$ & $0.20\leq C\leq 1.20$ & $0.40\leq C\leq
1.20$ \\ 
$A=0.013$ & $0.40\leq C\leq 1.20$ & $0.20\leq C\leq 1.20$ & $0.40\leq C\leq
1.20$ \\ \hline
\end{tabular}%
\end{center}
\caption{Coupling strengths in the range $0\leq C\leq 1.2$ for which the
responder is asymptotically stable. }
\label{Table asymp stability}
\end{table}

\item In general, when the noise amplitude increases, the threshold of
stable synchronization moves towards stronger couplings. However, the Model H%
\'{e}non 0.3-0.1 (Figure \ref{figure2}) shows that there can be parameter
settings for which that threshold is virtually the same for the noise
amplitudes considered here.
\end{itemize}

Finally, let us point out that we also performed numerical simulations with
perceptrons to detect the possible existence of the conventional
synchronization map (period 1). The results were similar but not as sharp
regarding weak synchronization states as the results obtained with LSTM nets
to detect synchronization maps of period $K>1$. However, no performance
analysis of Method 1 with respect to the period $K$ was carried out and,
hence, no attempt was made to optimize the parameter $K$ (which, anyway,
depends on the data at hand).


\section{Application to real-world data: EEGs}

\label{sec7}

The purpose of this section is to illustrate the application of Method 1 to
real world data, specifically, intracranial EEG recordings from a subject
with epilepsy. Therefore, we will not scrutinize here the complexity of such
signals but rather check whether our findings align with results obtained in
previous studies.

First of all, we notice that real observations $(\varphi _{X}(x_{t}))_{1\leq
t\leq T}$ and $(\varphi _{Y}(y_{t}))_{1\leq t\leq T}$ of coupled systems $X$
and $Y$, respectively, can deviate from our assumptions in Sections \ref%
{sec2}-\ref{sec6} in two important issues: nonstationarity or/and
bidirectionality of the coupling, as it actually occurs with the time series
in this section. To meet these challenges, this time we will apply Method 1
(Section \ref{sec6}) in both directions $X\rightsquigarrow Y$ and $%
Y\rightsquigarrow X$, on sufficiently short data segments to ensure
approximate stationarity.

There is a subtlety, though. In the unidirectionally coupling $%
X\rightsquigarrow Y$ studied in the previous sections, we detected
synchronization by detecting a functional dependency between $\mathbf{y}_{t}$
and $\mathbf{x}_{t},...,\mathbf{x}_{t-K+1}$, namely, $\mathbf{y}%
_{t}=H_{K,X\rightsquigarrow Y}(\mathbf{x}_{t},...,\mathbf{x}_{t-K+1})$,
where $H_{K,X\rightsquigarrow Y}$ is the reconstructed synchronization map
of period $K$ of the coupling $X\rightsquigarrow Y$, defined in equation (%
\ref{multi h reconst 2}). The robustness to noise of $H_{K,X\rightsquigarrow
Y}$ allowed us then to extend our conclusions to signals contaminated with
low-amplitude noise. If, for the sake of this argument, we think of a
bidirectional coupling $X\leftrightsquigarrow Y$ as the joint action of two
separate unidirectional couplings $X\rightsquigarrow Y$ and $%
Y\rightsquigarrow X$, then $\mathbf{y}_{t}$ will depend on $\mathbf{x}%
_{t},...,\mathbf{x}_{t-K+1}$ (whether $X$ and $Y$ are synchronized or not)
through the cross map of period $K$ of the coupling $Y\rightsquigarrow X$,
i.e., $\mathbf{y}_{t}=\Phi _{K,Y\rightsquigarrow X}(\mathbf{x}_{t},...,%
\mathbf{x}_{t-K+1})$; see equation (\ref{Cross(k) 4}) with $\mathbf{y}$ and $%
\mathbf{x}$ swapped. Therefore, here we expect $\mathbf{y}_{t}$ to depend on 
$\mathbf{x}_{t},...,\mathbf{x}_{t-K+1}$ in general.

The bottom line is that, by using Method 1 in the directions $%
X\rightsquigarrow Y$ and $Y\rightsquigarrow X$, we will be able to detect
the \textquotedblleft dominant driver\textquotedblright\ or the
\textquotedblleft coupling directionality\textquotedblright\ of the
bidirectional coupling $X\leftrightsquigarrow Y$.\ To this end, we are going
to measure the strength of the coupling in both directions via the accuracy
of the predictions of $x_{t}$\ and $y_{t}$\ made by LSTM nets in
short non-overlapping segments over the entire EEGs, the dominant driver
being given by the direction with the strongest coupling. In case of equal
strengths, the systems are assumed to be synchronized.


\subsection{Data description}

\label{sec7.1}

The data that we are going to analyze is the following; see Lehnertz and Dickten \cite%
{Lehnertz2015} for more detail.

\begin{enumerate}
\item The signals are EEGs recorded intracranially from a subject with
epilepsy during $86,090$ s $($23 h, 54 m, 50.4 s) with 48 electrode contacts
at a sampling frequency of 200 Hz (sampling time = 5 ms). The subject had
signed informed consent that her/his clinical data might be used and
published for research purposes, and the study protocol had previously been
approved by the ethics committee of the University of Bonn. The recording
started at 7:00 am, corresponding to the initial sampling interval $t=1$,
and ended at the final sampling time $t_{\mathrm{final}}=$ $17.217984\times
10^{6}$. The epileptic convulsions occurred at the following sampling times.

\begin{itemize}
\item Average time of a first group of subclinical seizures: $\bar{t}%
_{C1}=\allowbreak 3.\,\allowbreak 4082\times 10^{6}$ ($17,041$ s). By
subclinical seizures we mean localized seizure activity on the EEG with no
obvious clinical activity.

\item Average time of a second group of subclinical seizures: $\bar{t}%
_{C2}=4.\,\allowbreak 6082\times 10^{6}$ ($23,041$ s).

\item Average time of a third group of subclinical seizures: $\bar{t}%
_{C3}=6.\,\allowbreak 8762\times 10^{6}$ ($\allowbreak 34,381$ s).

\item Onset time of a clinical seizure (the only one in the whole series): $%
t_{C4}=17.1842\times 10^{6}$ ($85,921$ s).
\end{itemize}

\item A schematic of the implanted electrodes can be found in Figure 2
of Lehnertz and Dickten \cite{Lehnertz2015}. The electrodes contacts are divided into the following three
categories:

\begin{itemize}
\item focal (F), which comprises electrode contacts located within the
seizure-onset zone;

\item neighbor (N), which groups electrode contacs not more than two
contacts distant to those of category $\mathrm{F}$;

\item other (O), gathering all remaining electrode contacts.
\end{itemize}

To designate the electrode contacts and their (approximately) 24h recordings
we use the same labels as in \cite{Lehnertz2015}. For example, $X=\mathrm{%
TR01}$ means that the system $X$ is the source of the EEG $(\varphi
_{X}(x_{t}))_{1\leq t\leq t_{\mathrm{final}}}$ recorded at the electrode
contact \textrm{TR01}.

\item For the sake of our analysis, we will consider the following five
pairs $(X,Y)$ of electrode contacts.

\begin{itemize}
\item Case 1: $(X,Y)=(\mathrm{TR05}$-$\mathrm{TR06})$ in the categories
(F-F).

\item Case 2: $(X,Y)=(\mathrm{TR07}$-$\mathrm{TBPR1})$ in the categories
(F-N).

\item Case 3: $(X,Y)=(\mathrm{TR05}$-$\mathrm{TL05})$ in the categories
(F-O).

\item Case 4: $(X,Y)=(\mathrm{TBAR1}$-$\mathrm{TLL04})$ in the categories
(N-O).

\item Case 5: $(X,Y)=(\mathrm{TLR04}$-$\mathrm{TLL04})$ in the categories
(O-O).
\end{itemize}

\item As in the previous numerical simulations, the embedding dimension of
the systems $X$ and $Y$ is 5. Thus,%
\begin{equation}
\mathbf{x}_{t}=(\varphi _{X}(x_{t}),\varphi _{X}(x_{t+1}),...,\varphi _{X}(x_{t+4})), \label{vect}
\end{equation}%
$1\leq t\leq t_{\mathrm{final}}-4$, are the time-delay vectors corresponding
to system $X$, and analogously with the EEG $(\varphi _{Y}(y_{t}))_{1\leq
t\leq t_{\mathrm{final}}}$ generated by the system $Y$.

\item For approximate stationarity \cite{Rieke2002,Rieke2003}, we
partitioned the time series $(\mathbf{x}_{t})$ and $(\mathbf{y}_{t})$, $%
1\leq t\leq t_{\mathrm{final}}-4$, into 1434 non-overlapping segments 
\begin{equation}
S_{X,n}=(\mathbf{x}_{t})_{12000(n-1)+1\leq t\leq
12000n-4}, \label{S_X,n}
\end{equation}%
and 
\begin{equation}
S_{Y,n}=(\mathbf{y}_{t})_{12000(n-1)+1\leq t\leq
12000n-4},  \label{S_Y,n}
\end{equation}%
of 11,996 points ($\simeq 60$ s) each, $n=1,2,...,1434$, and a last pair of segments 
\begin{equation}
S_{X,1435}=(\mathbf{x}_{t})_{17208001\leq t\leq 17217980},  \label{S_X,1435}
\end{equation}%
and
\begin{equation}
S_{Y,1435}=(\mathbf{y}_{t})_{17208001\leq t\leq 17217980},  \label{S_Y,1435}
\end{equation}%
comprising only 9,980 points ($\simeq 50$ s). The segments $1\leq n\leq 720$%
, correspond to the daylight hours (7 am-7 pm), while the segments $721\leq
n\leq 1435$ correspond to the night hours. The clinical seizure occurs in
the segment $n=1433$, i.e., the third to last segment of the series, and it
initiates just one second after the beginning of that segment ($%
t_{C4}=85,921 $ s).

\item As in Section \ref{sec6}, we use the first $80\%$ of the data of each $%
n$th segment $S_{X,n}$ and $S_{Y,n}$ as training data, and the remaining $%
20\%$ as testing data. So, this time we obtain two accuracy measures: (i) $%
\mathrm{MSE}_{X\rightsquigarrow Y}(n)$, for the predictions of $\mathbf{y}%
_{t}$ output by the LSTM net, based on $\mathbf{x}_{t},...,\mathbf{x}%
_{t-K+1} $ with testing data of the segments $S_{Y,n}$ and $S_{X,n}$, and
(ii) $\mathrm{MSE}_{Y\rightsquigarrow X}(n)$, for the predictions of $%
\mathbf{x}_{t}$ output by the LSTM net, based on $\mathbf{y}_{t},...,\mathbf{%
y}_{t-K+1} $ with testing data of the segments $S_{X,n}$ and $S_{Y,n}$. As
in Section \ref{sec6}, we set $K=10$ here. Of course, the parameter $K$ can
be fine-tuned for optimal results, but this is an issue not contemplated in
the present work.
\end{enumerate}


\subsection{Results}

\label{sec7.2}

\begin{figure*}[tbh]
\begin{center}
\includegraphics[width=120mm]{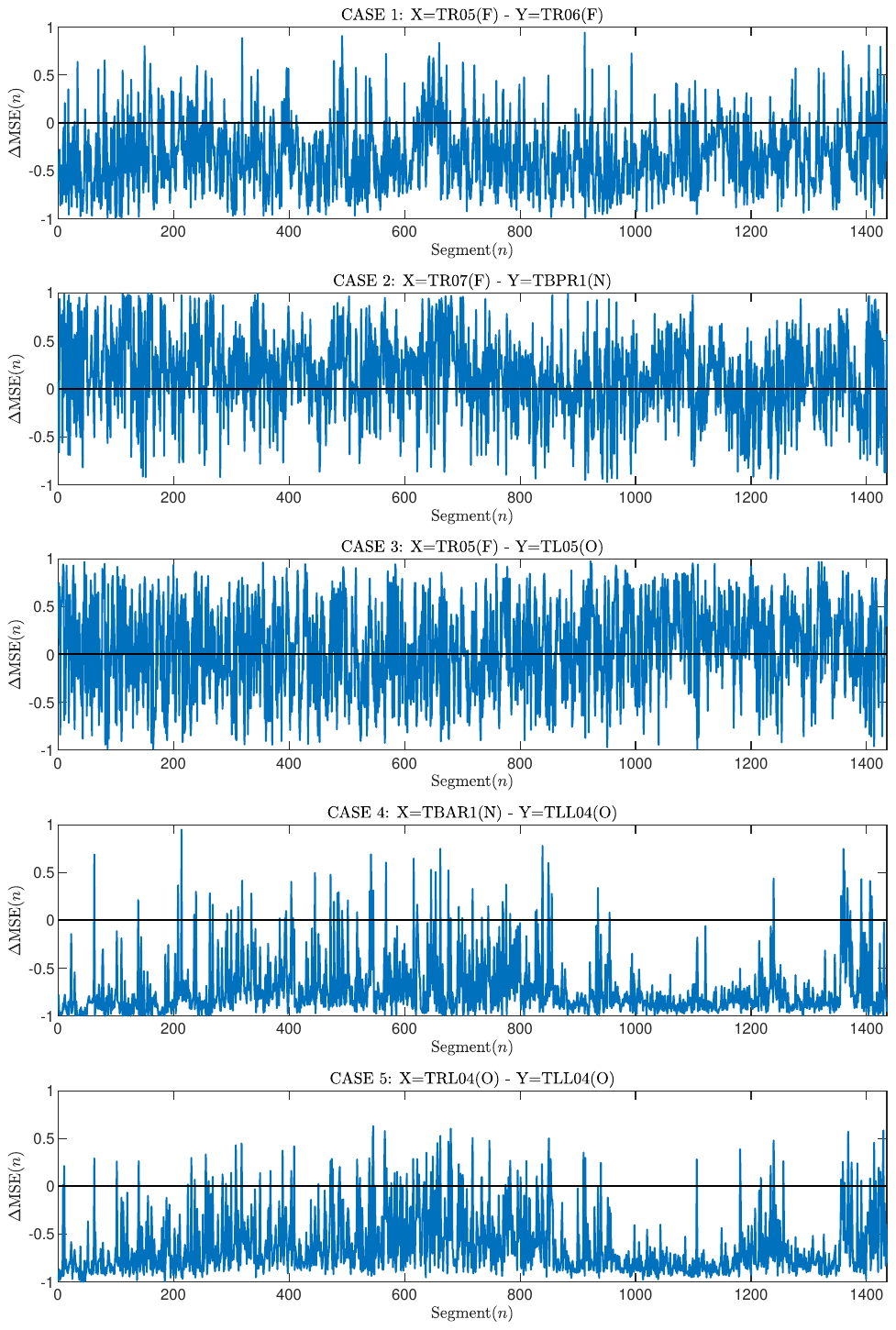}
\caption{ Top to bottom: plots of $\Delta \mathrm{MSE}(n)$, the
directionality indicator (\protect\ref{direct index}), obtained using the
segments $S_{X,n}$ and $S_{Y,n}$, $1\leq n\leq 1435$, given in equations (%
\protect\ref{S_X,n})-(\protect\ref{S_X,1435}), for Cases 1 to 5. The
clinical seizure occurs in the segment $n=1433$, too close to the right
margin to be marked. See Section \protect\ref{sec7.1} for detail.}
\label{figure4}
\end{center}
\end{figure*}

Since, at variance with the numerical simulations in Section \ref{sec6}, we
have here bidirectionally coupled signals and two prediction accuracy
measures $\mathrm{MSE}_{X\rightsquigarrow Y}(n)$ and $\mathrm{MSE}%
_{Y\rightsquigarrow X}(n)$, we are going to use the coupling directionality
index 
\begin{equation}
\Delta \mathrm{MSE}(n)=\frac{\mathrm{MSE}_{X\rightsquigarrow Y}(n)-\mathrm{%
MSE}_{Y\rightsquigarrow X}(n)}{\mathrm{MSE}_{X\rightsquigarrow Y}(n)+\mathrm{%
MSE}_{Y\rightsquigarrow X}(n)}  \label{direct index}
\end{equation}%
for each pair of data segments $S_{X,n}$ and $S_{Y,n}$, $1\leq n\leq 1435$,
so that

\begin{description}
\item[(i)] $-1\leq \Delta \mathrm{MSE}(n)\leq +1$, and

\item[(ii)] $\Delta \mathrm{MSE}(n)\geq 0$ if and only if $\mathrm{MSE}%
_{Y\rightsquigarrow X}(n)\leq \mathrm{MSE}_{X\rightsquigarrow Y}(n)$, i.e.,
knowledge of $\mathbf{x}_{t}$ in the $n$th segment leads to better
predictions of $\mathbf{y}_{t}$ than the other way around.
\end{description}

Therefore, if $\Delta \mathrm{MSE}(n)>0$ (resp., $\Delta \mathrm{MSE}(n)<0)$%
, then we conclude that $X$ is the dominant driver (resp., $Y$ is the
dominant driver). This interpretation agrees with other approaches based on
the cross map \cite{Sugihara2012,Amigo2018}, transfer entropy \cite%
{Schreiber2000,Staniek2008}, phase dynamics \cite{Rosenblum2001}, etc.
Otherwise, if $\Delta \mathrm{MSE}(n)=0$, then $X$ and $Y$ are assumed to be
synchronized in segment $n$ (although it might be difficult to discern this
situation from \textquotedblleft no-coupling\textquotedblright ).

Figure \ref{figure4} plots $\Delta \mathrm{MSE}(n)$ vs the segment number $n$%
, $1\leq n\leq 1435$. The numerical results are summarized in Table \ref%
{Table24h} for the 24h EEGs, and in Table \ref{Table12h} for 12h EEGs
corresponding to daylight hours ($1\leq n\leq 720$) and night hours ($%
721\leq n\leq 1435$). It was not possible to highlight the clinical seizure
in Figure \ref{figure4} because it occurs in the segment $n=1433$, so any
visual marks at that point are indistinguishable from the right margin of
the corresponding panel.

\begin{table}[tbh]
\par
\begin{center}
\begin{tabular}{|c|c|c|}
\hline
\textbf{Case} & $\Delta \mathrm{MSE}(n)>0$ & \textbf{Dominant electrode} \\ 
\hline
1 (F-F) & $16\%$ of segments & $Y=$ $\mathrm{TR06}$ (F) dominates $X=\mathrm{%
TR05}$ (F) \\ \hline
2 (F-N) & $64\%$ of segments & $X=$ $\mathrm{TR07}$ (F) dominates $Y=\mathrm{%
TBPR1}$ (N) \\ \hline
3 (F-O) & $56\%$ of segments & $X=$ $\mathrm{TR05}$ (F) dominates $Y=\mathrm{%
TL05}$ (O) \\ \hline
4 (N-O) & $5\%$ of segments & $Y=$ $\mathrm{TLL04}$ (O) dominates $X=\mathrm{%
TBAR1}$ (N) \\ \hline
5 (O-O) & $7\%$ of segments) & $Y=$ $\mathrm{TLL04} $ (O) dominates $X=%
\mathrm{TLR04}$ (O) \\ \hline
\end{tabular}%
\end{center}
\caption{Results of Cases 1-5 with 24h EEGs. }
\label{Table24h}
\end{table}

\begin{table}[tbh]
\par
\begin{center}
\begin{tabular}{|c|c|c|}
\hline
\textbf{Case} & $\Delta \mathrm{MSE}(n)>0$ \textbf{day} & $\Delta \mathrm{MSE%
}(n)>0$ \textbf{night} \\ \hline
1 (F-F) & $18\%$ ($Y$ dominant) & $15\%$ ($Y$ dominant) \\ \hline
2 (F-N) & $70\%$ ($X$ dominant) & $58\%$ ($X$ dominant) \\ \hline
3 (F-O) & $51\%$ ($X$ dominant) & $60\%$ ($X$ dominant) \\ \hline
4 (N-O) & $6\%$ ($Y$ dominant) & $4\%$ ($Y$ dominant) \\ \hline
5 (O-O) & $9\%$ ($Y$ dominant) & $6\%$ ($Y$ dominant) \\ \hline
\end{tabular}%
\end{center}
\caption{Results of Cases 1-5 with 12h EEGs (day and night). }
\label{Table12h}
\end{table}

In view of Figure \ref{figure4} and Tables \ref{Table24h} and \ref{Table12h}%
, we can draw the following general conclusions.

\begin{itemize}
\item The coupling directionality, as measured by $\Delta \mathrm{MSE}(n)$,
depends on the segment $n$. The overall dominance of the signals is stable
with respect to day and night, although the \textit{dominance degrees}, as
measured by the percentages of segments contributing to the dominant
coupling direction in the first or second 12 hours, respectively, are
different in all cases.

\item Except in Case 2, the sign of $\Delta \mathrm{MSE}(n)$ during the
epileptic convulsions ($n=1433$) coincides with the overall coupling
direction. In fact, 
\begin{equation*}
\begin{tabular}{|l|c|c|c|c|c|}
\hline
\textbf{Case} & 1 & 2 & 3 & 4 & 5 \\ \hline
$\Delta \mathrm{MSE}(1433)$ & $-0.58$ & $-0.87$ & $0.60$ & $-0.68$ & 
$-0.40$ \\ \hline
\end{tabular}%
\end{equation*}

\item According to Osterhage et al. \cite{Osterhage2007,Osterhage2008}, an important question
in epileptology is whether the pathological interaction between the
seizure-onset zone (label F) and other brain areas (labels N and O), a
phenomenon called \textit{focal driving}, can also be identified during
seizure-free periods. Cases 2 and 3 in Table \ref{Table24h} answer this
question affirmatively, that is, our method detects focal driving in the
analyzed EEG.

\item In addition, Cases 2 and 3 in Table \ref{Table12h} indicate that focal
driving is not diminished during sleep.

\item More generally, Table \ref{Table12h} shows that focal driving does not
appear to be influenced by other (possibly \textquotedblleft
stronger\textquotedblright ) synchronization phenomena such as sleep.

\item The dominance degree is rather high in the Cases 1, 4 and 5, with $%
\Delta \mathrm{MSE}(n)<0$ over 80\% of the segments both in the 24h and 12h
EEGs. Note that the interaction in those cases is local (Cases 1 and 5) or
it does not involve the seizure generating area (Case 4).
\end{itemize}

The above findings are in line with the results of more comprehensive
studies by Lehnertz and Dickten \cite{Lehnertz2015}, Dickten et al. \cite{Dickten2016}, 
and Osterhage et al. \cite{Osterhage2007,Osterhage2008}, which empirically demostrates the capability of our
LSTM net-based method.


\section{Conclusion}

\label{sec8}

Synchronization of two unidirectionally coupled dynamical systems $%
X\rightsquigarrow Y$ is a classical topic in nonlinear dynamics. It is
defined by the existence of a continuous function $y_{t}=h(x_{t})$ between
the states $x_{t}$ of the driver $X$ and the states $y_{t}$ of the
responder, called the synchronization map. While $h$, when it exists, points
from the state space of the driver (domain) to the state space of the
responder (range), the cross map $\mathbf{x}_{t}=\Phi (\mathbf{y}_{t})$
always exists in that framework, is continuous, and points in the opposite
direction between the corresponding reconstructed state spaces. In the
standard, noiseless scenario, the existence of the synchronization map
(i.e., synchronization between $X$ and $Y$) amounts to $\Phi $ being
invertible and bicontinuous. These and other fundamentals of generalized
synchronization in the absence of dynamical noise were presented in a
self-contained and unified way in Section \ref{sec2}, with emphasis on the
relationship between the cross map and the synchronization map.

In this context, the main contributions of the present paper are the
following.

(1) \textit{Introduction of higher-period versions of the cross and
synchronization maps} in equations (\ref{Cross(k) 4}) and (\ref{multiv h}),
the period-1 versions corresponding to the conventional concepts. They are
based on the corresponding maps of order $k$, defined in equations (\ref%
{Cross(k) 1}) and (\ref{sync map B}), and, actually, they may be defined in
many different ways. A synchronization map of period $10$ was used in the
numerical simulation of Section \ref{sec6} because it gave better results
than the conventional map in the detection of synchronization. Higher-period
cross maps were invoked in Section \ref{sec7} to understand the sign of the
directionality index (\ref{direct index}) when the coupling is
bidirectional. Optimization of the period was not discussed because it is
beyond the scope of this paper.

(2) \textit{Generalizations of the synchronization map and the cross map
when the driver is noisy} in Sections \ref{sec5.1} and \ref{sec5.2},
respectively. To this end, the dynamical noise was modeled as stochastic
forcing. The generalizations consist of families of maps that depend on
noise parameters and coincide with their conventional counterparts when the
noise is switched off. As usual, those generalizations have the wished
properties under some formal provisos, e.g., generacy of the maps and
parameters involved, as well as the driver states in the case of the cross
map. But this does not mean that they are only useful in theory; they can be
also useful in practice, e.g., in laboratory or numerical experiments, where
typical properties are taken for granted and even noise parameters may be
known.

(3) \textit{Application of LSTM nets to detect synchronization in synthetic
data}. This method harness the existence of synchronization maps of higher
periods in both the noiseless and noisy scenarios. To be more precise, in
the numerical simulations of Section \ref{sec6}, synchronization was
revealed in the reconstructed state spaces by detecting a period-10
synchronization map (equation (\ref{multi h reconst 2})) using an LSTM net
and predictability. The dynamical systems were two coupled H\'{e}non maps
with several parameter settings and noise amplitudes. As a benchmark we used
nearest-neighbor cross prediction based on the continuity of the cross map
and its inverse (in case of synchronization). The agreement of the results
obtained with the two methods was excellent. The results also showed the
robustness of both methods against noise.

(4) \textit{Application of LSTM nets to detect the dominant driver in real
world data} in Section \ref{sec7}. The real world data consisted of a 24h
EEG from a subject with epilepsy. To cope with the nonstationarity of the
signal and the bidirectionality of the couplings between different brain
areas, we partitioned the EEG in non-overlapping 60s segments and measured
the coupling dominance by the directionality index $\Delta \mathrm{MSE}(n)$
defined in equation (\ref{direct index}); this index is based on the mean
square error of predictions made by LSTM nets in the $n$th segment. The
results agreed with results published in the literature, in particular, the
existence of focal driving and its robustness to the day/night cycle.

\section{Acknowledgments}

J.M. Amig\'{o} and R. Dale were financially supported by Agencia Estatal de
Investigaci\'{o}n, Spain, grant
PID2019-108654GB-I00/AEI/10.13039/501100011033. J.M. Amig\'{o} was also
supported by Generalitat Valenciana, Spain, grant PROMETEO/2021/063.

\bigskip \bigskip

\noindent DATA AVAILABILITY STATEMENT

The EEG dataset presented in this article is not readily available because
it contains information that could compromise the privacy of the research
participant. Requests to access the dataset should be directed to
klaus.lehnertz@ukbonn.de.

\bigskip \bigskip

\noindent ETHICS STATEMENT

The studies involving human participants were reviewed and approved by the
ethics committee of the University of Bonn. The subject provided written
informed consent to participate in this study.

\bigskip



\begin{thebibliography}{99}
\bibitem{Afraimovich1986} V.S. Afraimovich, N.N. Verichev, and M.I.
Rabinovich, \textquotedblleft Stochastic synchronization of oscillation in
dissipative systems,\textquotedblright\ Radiophysics and Quantum Electronics 
\textbf{29}, 795-803 (1986).

\bibitem{Rulkov1995} N.F. Rulkov, M.M. Sushchik, L.S. Tsimring, and H.D.I.
Abarbanel, \textquotedblleft Generalized synchronization of chaos in
directionally coupled chaotic systems,\textquotedblright\ Physical Review E 
\textbf{51}, 980-994 (1995).

\bibitem{Rosenblum1997} M.G. Rosenblum, A.S. Pikovsky, and J. Kurths,
\textquotedblleft From phase to lag synchronization in coupled chaotic
oscillators,\textquotedblright\ Physical Review Letters \textbf{78},
4193-4196 (1997).

\bibitem{Pecora1990} L.M. Pecora and T.L. Carroll, \textquotedblleft
Synchronization in chaotic systems,\textquotedblright\ Physical Review
Letters \textbf{64}, 821-824 (1990).

\bibitem{Boccaletti2002} S. Boccaletti, J. Kurths, G. Osipov, D.L.
Valladare, and C.S. Zhou, \textquotedblleft The synchronization of chaotic
systems,\textquotedblright\ Physics Reports \textbf{366}, 1-101 (2002).

\bibitem{Cuomo1993} K.M. Cuomo, A.V. Oppenheim, and S.H. Strogatz,
\textquotedblleft Synchronization of Lorenz-Based Chaotic Circuits with
Applications to Communications,\textquotedblright\ IEEE Transactions on
Circuits and Systems II \textbf{40}, 626-633(1993).

\bibitem{Kocarev1995} L. Kocarev and U. Parlitz, \textquotedblleft General
approach for chaotic synchronization with applications to
communication,\textquotedblright\ Physical Review Letters \textbf{74},
5028-5031 (1995).

\bibitem{Nowotny2008} T. Nowotny, R. Huerta, and M.I. Rabinovich,
\textquotedblleft Neuronal synchrony: Peculiarity and
generality,\textquotedblright\ Chaos 18, 037119 (2008).

\bibitem{Amigo2015} J.M. Amig\'{o}, T.S. Mosqueiro, and R. Huerta,
\textquotedblleft Predicting synchronization of three mutually inhibiting
groups of oscillators with strong resetting,\textquotedblright\ Applied
Mathematics and Information Sciences \textbf{9}, 2245-2256 (2015) .

\bibitem{Hovel2020} P. H\"{o}vel, A. Viol, P. Loske. L. Merfort, and V.
Vuksanovi\'{c}, \textquotedblleft Synchronization in functional networks of
the human brain,\textquotedblright\ Journal of Nonlinear Science 30,
2259--2282 (2020).

\bibitem{Kocarev1992} L.J. Kocarev, K.S. Halle, K. Eckert, L.O. Chua, and U.
Parlitz, \textquotedblleft Experimental demonstration of secure
communications via chaotic synchronization,\textquotedblright\ International
Journal of Bifurcation and Chaos 2, 709-713 (1992).

\bibitem{Amigo2007} J.M. Amig\'{o}, L. Kocarev, and J. Szczepanski,
\textquotedblleft Discrete Lyapunov exponent and resistance to differential
cryptanalysis,\textquotedblright\ IEEE Transactions on Circuits and Systems
II \textbf{54}, 882-886 (2007).

\bibitem{Kinzel2010} W. Kinzel, A. Englert, and I. Kanter, \textquotedblleft
On chaos synchronization and secure communication,\textquotedblright\
Philosophical Transactions of the Royal Society A \textbf{368}, 379--389
(2010).

\bibitem{Banerjee2011} S. Banerjee (Editor), \textit{Chaos Synchronization
and Cryptography for Secure Communications: Applications for Encryption},
Information Science Reference, Hershey PA (USA) 2011.

\bibitem{Pikovsky2001} A. Pikovsky, M. Rosenblum, and J. Kurths, \textit{%
Synchronization: A Universal Concept in Nonlinear Sciences}, Cambridge
University Press, Cambridge, UK, 2001.

\bibitem{Pecora2015} L.M. Pecora and T.L. Carroll, \textquotedblleft
Synchronization of chaotic systems,\textquotedblright\ Chaos \textbf{25},
097611 (2015).

\bibitem{Schiff1996} S.J. Schiff, P. So, T. Chang, R.E. Burke, and T. Sauer,
\textquotedblleft Detecting dynamical interdependence and generalized
synchrony through mutual prediction in a neural ensemble,\textquotedblright\
Physical Review E \textbf{54}, 6708-6724 (1996).

\bibitem{LeVanQuyen1999} M. Le Van Quyen, J. Martinerie, C. Adam, and F.J.
Varela, \textquotedblleft Nonlinear analyses of interictal EEG map the brain
interdependences in human focal epilepsy,\textquotedblright\ Physica D 
\textbf{127}, 250-266 (1999).

\bibitem{Hunt1997} B.R. Hunt, E. Ott, and J.A. Yorke, \textquotedblleft
Differentiable generalized synchronization of chaos,\textquotedblright\
Physical Review E \textbf{55}, 4029-4034 (1997).

\bibitem{Pyragas1997} K. Pyragas, \textquotedblleft Conditional Lyapunov
exponents from time series,\textquotedblright\ Physical Review E \textbf{56}%
, 5183-5188 (1997).

\bibitem{Abarbanel1996} H.D.I. Abarbanel, N.F. Rulkov, and M.M. Sushchick,
\textquotedblleft Generalized synchronization of chaos: the auxiliary system
approach,\textquotedblright\ Physical Review E \textbf{53}, 4528-4535 (1996).

\bibitem{Kocarev1996} L. Kocarev and U. Parlitz, \textquotedblleft
Generalized synchronization, predictability and equivalence of
unidirectionally coupled dynamical systems,\textquotedblright\ Physical
Review Letter \textbf{76}, 1816-1819 (1996).

\bibitem{Arnhold1999} J. Arnhold, P. Grassberger, K. Lehnertz, and C.E.
Elger, \textquotedblleft A robust method for detecting interdependences:
application to intracranially recorded EEG,\textquotedblright\ Physica D 
\textbf{134}, 419-430 (1999).

\bibitem{QuianQuiroga2000} R. Quian Quiroga, J. Arnhold, and P. Grassberger,
\textquotedblleft Learning driver-response relationships from
synchronization patterns,\textquotedblright\ Physical Review E \textbf{61},
5142-5148 (2000).

\bibitem{QuianQuiroga2002} R. Quian Quiroga, A. Kraskov, T. Kreuz, and P.
Grassberger, \textquotedblleft Performance of different synchronization
measures in real data: A case study on electroencephalographic
signals,\textquotedblright\ Physical Review E \textbf{65}, 041903 (2002).

\bibitem{Sowa2005} R. Sowa, A. Chernihovskyi, F. Mormann, and K. Lehnertz,
\textquotedblleft Estimating phase synchronization in dynamical systems
using cellular nonlinear networks,\textquotedblright\ Physical Review E 
\textbf{71}, 061926 (2005).

\bibitem{Krug2007} D. Krug, H. Osterhage, C.E. Elger, and K. Lehnertz,
\textquotedblleft Estimating nonlinear interdependencies in dynamical
systems using cellular nonlinear networks,\textquotedblright\ Physical
Review E \textbf{76}, 041916 (2007).

\bibitem{Monetti2009} R. Monetti, W. Bunk, T. Aschenbrenner, and F.
Jamitzky, \textquotedblleft Characterizing synchronization in time series
using information measures extracted from symbolic
representations,\textquotedblright\ Physical Review E \textbf{79}, 046207
(2009).

\bibitem{Amigo2012} J.M. Amig\'{o}, R. Monetti, T. Aschenbrenner, and W.
Bunk, \textquotedblleft Transcripts: An algebraic approach to coupled time
series,\textquotedblright\ Chaos \textbf{22}, 013105 (2012).

\bibitem{Amigo2013} J.M. Amig\'{o} and K. Keller, \textquotedblleft
Permutation entropy: One concept, two approaches,\textquotedblright\
European Physical Journal Special Topics \textbf{222}, 263-274 (2013).

\bibitem{IbanezSoria2018} D. Ib\'{a}\~{n}ez-Soria, J. Garc\'{\i}a-Ojalvo, A.
Soria-Frisch, and G. Ruffini, \textquotedblleft Detection of generalized
synchronization using echo state networks,\textquotedblright\ Chaos \textbf{%
28}, 033118 (2018).

\bibitem{Lymburn2019} T. Lymburn, D.M. Walker, M. Small, and T. J\"{u}%
ngling, \textquotedblleft The reservoir's perspective on generalized
synchronization,\textquotedblright\ Chaos \textbf{29}, 093133 (2019).

\bibitem{Kantz2004} H. Kantz and T. Schreiber, \textit{Nonlinear Time Series
Analysis} (2nd edition), Cambridge University Press, Cambridge, UK, 2004.

\bibitem{Amigo2018} J.M. Amig\'{o} and Y. Hirata, \textquotedblleft
Detecting directional couplings from multivariate flows by the joint
distance distribution,\textquotedblright\ Chaos \textbf{28}, 075302 (2018).

\bibitem{Sugihara2012} G. Sugihara, R. May, H. Ye, C.-H. Hsieh, E. Deyle, M.
Fogarty, and S. Munch, \textquotedblleft Detecting causality in complex
ecosystems,\textquotedblright\ Science \textbf{338}, 496-500 (2012).

\bibitem{Harnack2017} D. Harnack, E. Laminski, M. Sch\"{u}nemann, and K.R.
Pawelzik, \textquotedblleft Topological causality in dynamical
systems,\textquotedblright\ Physical Review Letters \textbf{119}, 098301
(2017).

\bibitem{Ying2022} X. Ying, S.-Y. Leng, H.-F. Ma, Q. Nie, Y.-C. Lai, and W.
Lin, \textquotedblleft Continuity scaling: A rigorous framework for
detecting and quantifying causality accurately,\textquotedblright\ Research 
\textbf{2022}, 9870149 (2022).

\bibitem{Muldoon1998} M.R. Muldoon, D.S. Broomhead, J.P. Huke, and R.
Hegger, \textquotedblleft Delay embedding in the presence of dynamical
noise,\textquotedblright\ Dynamics and Stability of Systems \textbf{13},
175--186 (1998).

\bibitem{Stark2003} J. Stark, D.S. Broomhead, M.E. Davies, and J. Huke,
\textquotedblleft Delay embeddings for forced systems. II: Stochastic
forcing,\textquotedblright\ Journal of Nonlinear Science \textbf{13},
519-577 (2003).

\bibitem{Stark1999} J. Stark, \textquotedblleft Delay embeddings for forced
systems. I: Deterministic forcing,\textquotedblright\ Journal of Nonlinear
Science \textbf{9}, 255-332 (1999).

\bibitem{Goodfellow2016} I. Goodfellow, Y. Bengio, and A. Courville, \textit{%
Deep Learning}, The MIT Press, Cambridge MA, 2016.

\bibitem{Takens1981} F. Takens, \textquotedblleft Detecting strange
attractors in turbulence,\textquotedblright\ in \textit{Dynamical Systems
and Turbulence}, Lecture Notes in Mathematics 898, edited by D.A. Rand and
L.S. Young (Springer Verlag, 1981), pp. 366-381.

\bibitem{Sauer1991} T. Sauer, J.A. Yorke, and M. Casdagli, \textquotedblleft
Embedology,\textquotedblright\ Journal of Statistical Physics \textbf{65},
579--616 (1991).

\bibitem{Whitney1936} H. Whitney, \textquotedblleft Differentiable
manifolds,\textquotedblright\ Annals of Mathematics \textbf{37}, 645-680
(1936).

\bibitem{Kennel1992} M.B. Kennel, R. Brown, and H.D.I. Abarbanel,
\textquotedblleft Determining embedding dimension for phase-space
reconstruction using a geometrical construction,\textquotedblright\ Physival
Review A \textbf{45}, 3403--3411 (1992).

\bibitem{Parlitz2012} U. Parlitz, \textquotedblleft Detecting generalized
synchronization,\textquotedblright\ Nonlinear Theoy and Applications, IEICE, 
\textbf{3}, 113-127 (2012).

\bibitem{Parlitz1997} U. Parlitz, L. Junge, and L. Kocarev,
\textquotedblleft Subharmonic entrainment of unstable period orbits and
generalized synchronization,\textquotedblright\ Physical Review Letters 
\textbf{79}, 3158--3161 (1997).

\bibitem{Pecora1995} L.M. Pecora, T.L. Carroll, and J.F. Heagy,
\textquotedblleft Statistics for mathematical properties of maps between
time series embeddings,\textquotedblright\ Physical Review E \textbf{52},
3420--3439 (1995).

\bibitem{Rulkov2001} N.F. Rulkov, V.S. Afraimovich, C.T. Lewis, J.R.
Chazottes, and A. Cordonet, \textquotedblleft Multivalued mappings in
generalized chaos synchronization,\textquotedblright\ Physical Review E 
\textbf{64}, 016217 (2001).

\bibitem{Amigo2013B} J.M. Amig\'{o}, P. Kloeden, and A. Gim\'{e}nez,
\textquotedblleft Switching systems and entropy,\textquotedblright\ Journal
of Difference Equations and Applications \textbf{19}, 1872-1888 (2013).

\bibitem{Amigo2013C} J. M. Amig\'{o}, P. Kloeden, and A. Gim\'{e}nez,
Entropy increase in switching systems,\ Entropy \textbf{15}, 2363--2383
(2013).

\bibitem{Barnsley1985} M. F. Barnsley and S. Demko, \textquotedblleft
Iterated function systems and the global construction of
fractals,\textquotedblright\ Proceedings of the Royal Society London A 
\textbf{339}, 243-375 (1985).

\bibitem{Kloeden2011} P.E. Kloeden and M. Rasmusen, \textit{Nonautonomous
Dynamical Systems}, American Mathematical Society, Providence (Rhode
Island), 2011.

\bibitem{Arnold2003} L. Arnold, \textit{Random Dynamical Systems}, Springer
Verlar, Berlin, 2003.

\bibitem{Hirata2021} Y. Hirata, J.M. Amig\'{o}, S. Horai, K. Ogimoto, and K.
Aihara, \textquotedblleft Forecasting wind power ramps with prediction
coordinates,\textquotedblright\ Chaos \textbf{31}, 103105 (2021).

\bibitem{Petersen2000} K. Petersen, \textit{Ergodic Theory}. Cambridge
Unversity Press, Cambridge UK, 2000.

\bibitem{Moskalenko2011} O.I. Moskalenko, A. E. Hramov, A.A. Koronovskii,
and A.A. Ovchinnikov, \textquotedblleft Effect of noise on generalized
synchronization of chaos: theory and experiment,\textquotedblright\ European
Physical Journal B \textbf{82}, 69--82 (2022).

\bibitem{Monster2017} D. M\o nster, R. Fusaroli, K. Tyl\'{e}n, A. Roepstorff
and J.F. Sherson, \textquotedblleft Causal inference from noisy time-series
data -- Testing the Convergent Cross-Mapping algorithm in the presence of
noise and external influence,\textquotedblright\ Future Generation Computer
Systems \textbf{73}, 52-62 (2017).

\bibitem{Datseris2022} G. Datseris and U. Parlitz, \textit{Nonlinear Dynamics%
}, Springer Nature 2022.

\bibitem{Lehnertz2015} K. Lehnertz and H. Dickten, \textquotedblleft
Assesing directionality and strength of coupling through symbolic analysis:
an application to epilepsy patients,\textquotedblright\ Philosophical
Transactions of the Royal Society A \textbf{373}, 20140094 (2015).

\bibitem{Dickten2016} H. Dickten. S. Porz, C.E. Elger and K. Lehnertz,
\textquotedblleft Weighted and directed interactions in evolving large-scale
epileptic brain networks,\textquotedblright\ Scientific Reports \textbf{6},
34824 (2016).

\bibitem{Osterhage2007} H. Osterhage, F. Mormann, T. Wagner, and K.
Lehnertz, \textquotedblleft Measuring the directionality of coupling: Phase
versus state space dynamics and applications to EEG time
series,\textquotedblright\ International Journal of Neural Systems \textbf{17%
}, 139--148 (2007).

\bibitem{Osterhage2008} H. Osterhage, F. Mormann, T. Wagner, and K.
Lehnertz, \textquotedblleft Detecting directional coupling in the human
epileptic brain: Limitations and potential pitfalls,\textquotedblright\
Physical Review E \textbf{77}, 011914 (2008).

\bibitem{Rieke2002} C. Rieke, K. Sternickel, R.G. Andrzejak, C.E. Elger, P.
David, and K. Lehnertz, \textquotedblleft Measuring nonstationarity by
analyzing the loss of recurrence in dynamical systems,\textquotedblright\
Physical Review Letters \textbf{88}, 244102 (2002).

\bibitem{Rieke2003} C. Rieke, F. Mormann, R.G. Andrzejak, T. Kreuz, P.
David, C.E. Elger, and K. Lehnertz, \textquotedblleft Discerning
nonstationarity from nonlinearity in seizure-free and preseizure EEG
recordings from epilepsy patients,\textquotedblright\ IEEE Transactions on
Biomedical Engineering \textbf{50}, 634--639 (2003).

\bibitem{Schreiber2000} T. Schreiber, \textquotedblleft Measuring
information transfer,\textquotedblright\ Physical Review Letters \textbf{85}%
, 461--464 (2000).

\bibitem{Staniek2008} M. Staniek and K. Lehnertz, \textquotedblleft Symbolic
transfer entropy,\textquotedblright\ Physical Review Letters \textbf{100},
(2008) 158101.

\bibitem{Rosenblum2001} M.G. Rosenblum and A. Pikovsky, \textquotedblleft
Detecting direction of coupling in interacting
oscillators,\textquotedblright\ Physical Review E \textbf{64}, 045202 (2001).
\end{thebibliography}
\end{document}